# Large Enhancements in Optical and Piezoelectric Properties in Ferroelectric $Zn_{1-x}Mg_xO$ Thin Films through Engineering Electronic and Ionic Anharmonicities

*Rui Zu, Gyunghyun Ryu, Kyle P. Kelley, Steven M. Baksa, Leonard C Jacques, Bo Wang, Kevin Ferri, Jingyang He, Long-Qing Chen, Ismaila Dabo, Susan Trolier-McKinstry, Jon-Paul Maria, Venkatraman Gopalan**

R. Zu, G. Ryu, S.M. Baksa, L.C. Jacques, B. Wang, K. Ferri, J. He, L.Q. Chen, I. Dabo, S. Trolier-McKinstry, J-P. Maria, V. Gopalan
Department of Materials Science and Engineering, The Pennsylvania State University, University Park, Pennsylvania, 16802, USA
E-mail: vxg8@psu.edu

K.P. Kelly
Center for Nanophase Materials Sciences, Oak Ridge National Laboratory, Oak Ridge, TN 37831, USA

L.Q. Chen, V. Gopalan
Department of Engineering Science and Mechanics, The Pennsylvania State University, University Park, Pennsylvania 16802, USA

L.Q. Chen
Department of Mathematics, The Pennsylvania State University, University Park, Pennsylvania 16802, USA

S. Trolier-McKinstry, V. Gopalan
Materials Research Institute, The Pennsylvania State University, University Park, Pennsylvania 16802, USA

V. Gopalan
Department of Physics, Pennsylvania State University, University Park, Pennsylvania, 16802, USA

**Abstract**

Multifunctionality as a paradigm requires materials exhibiting multiple superior properties. Integrating second-order optical nonlinearity and large bandgap with piezoelectricity could, for example, enable broadband, strain-tunable photonics. Though very different phenomena at distinct frequencies, both second-order optical nonlinearity and piezoelectricity are third-rank polar tensors present only in acentric crystal structures. However, simultaneously enhancing both phenomena is highly challenging since it involves competing effects with tradeoffs. Recently, a large switchable ferroelectric polarization of ~ 80 $\mu C \cdot cm^{-2}$ was reported in $Zn_{1-x}Mg_xO$ films. Here, ferroelectric $Zn_{1-x}Mg_xO$ is demonstrated to be a platform that hosts simultaneously a 30% increase in the electronic bandgap, a 50% enhancement in the second harmonic generation coefficients, and a near 200% improvement in the piezoelectric coefficients over pure ZnO. These enhancements are shown to be due to a 400% increase in the electronic anharmonicity and a ~200% decrease in the ionic anharmonicity with Mg substitution. Precisely controllable periodic ferroelectric domain gratings are demonstrated down to 800 nm domain width, enabling ultraviolet quasi-phase-matched optical harmonic generation as well as domain-engineered piezoelectric devices.

## 1. Introduction

The idea of fully integrated strain-tunable optical devices such as adaptive optics, strain tunable photonic crystals and phase-matched nonlinear optical conversion devices requires the integration of piezoelectric and optical materials.[1–8] A single multifunctional material that could achieve both effective strain tuning of materials properties and exceptional optical response would be ideal from an engineering design perspective, but relatively few candidates are known, and of these, many are limited to bulk single crystals.[9,10]

Both piezoelectricity and optical second harmonic generation (SHG), the subjects of this study, are third-rank tensor properties with the same tensor form; both require acentric structures with broken spatial inversion symmetry. Since the most direct way to break inversion symmetry is in a polar structure with a spontaneous polarization, $P_s$, many of the well-known piezoelectric and SHG crystals are also polar and often ferroelectric. Ferroelectricity induces an anharmonic ionic potential well and, in turn, an anharmonic electronic potential well. Therefore, conventional wisdom suggests that enhancing $P_s$ would favor both piezoelectric and SHG tensors, though the magnitudes of the coefficients depend not just on the polarization, but also the polarizability. This picture however is oversimplified. It is often possible to enhance the piezoelectricity by increasing the dielectric constant, and in many cases, this is coupled with a concomitant decrease in the spontaneous polarization, for example, in the wurtzite structures[9] as well as in numerous normal and relaxor ferroelectric perovskites.[11,12] Further, although molecules with a larger electric dipole,[13,14] and crystals with larger $P_s$,[15–19] often exhibit larger SHG responses, many commercial nonlinear optical crystals belong to the nonpolar point group $\bar{4}2m$.[20] It is shown in this work that instead of spontaneous polarization, the anharmonicities of the electronic and ionic potential wells determine the SHG and piezoelectric properties. For optical frequency conversion such as SHG, a broader frequency range of optical transparency is desired, which requires a larger electronic bandgap, $E_g$. However, a larger bandgap is known to reduce the SHG coefficients dramatically, limiting high conversion efficiency in the ultraviolet range.[20,21] Thus, increasing the piezoelectric tensor, $d_{ijk}^{\text{Piezo}}$, the SHG tensor, $d_{ijk}^{\text{SHG}}$, and the electronic bandgap, $E_g$, simultaneously in one material is a challenging balancing act.

Nonetheless, each of these phenomena, namely the $d_{ijk}^{\text{Piezo}}$,[22–24] the $d_{ijk}^{\text{SHG}}$,[25,26] the bandgap,[27–29] and in addition, the electro-optic tensor,[30] have been separately reported to be enhanced with Mg substitution in ZnO. This work revisits and confirms the first three of these experimental measurements in a single set of films, though the magnitudes of enhancement reported here significantly differ from the literature for $d_{ijk}^{\text{SHG}}$. Importantly, this work presents a unified theoretical framework for understanding these counterintuitive simultaneous enhancements. Finally, it reports for the first time, precise ferroelectric domain engineering on the sub-micrometer scale.

ZnO is a well-known wide-bandgap semiconductor ($E_g \sim 3.2$ eV). Recently, $Zn_{1-x}Mg_xO$ was shown to have a giant switchable spontaneous polarization of ~ 80 μC/cm$^2$.[31] In this work, it is demonstrated that the electronic bandgap, the SHG coefficient, and the piezoelectric coefficient are simultaneously enhanced by Mg substitution in $Zn_{1-x}Mg_xO$. Mg substitution induces a 30% increase in the bandgap and a simultaneous 50% enhancement in the SHG coefficients, contrary to the well-known inverse relationship between the two properties in the literature.[20,21] This is shown to be due to a 400% increase in the electronic anharmonicity that offsets the increasing bandgap and results in a net increase in the $d^{\text{SHG}}$. The ferroelectric polarization $P_s$ decreases with higher Mg concentration (i.e., it anticorrelates to the increasing $d^{\text{SHG}}$), in contrast to the reported proportionality between $P_s$ and SHG response.[15,16] Near 200% enhancement in the piezoelectric coefficient is observed in $Zn_{1-x}Mg_xO$ with increasing Mg addition, which is attributed to an increase in the low-frequency dielectric constant arising in part due to the softening of the wurtzite structure (reduced *c*/*a* ratio)[32,33] and a decrease in the anharmonicity of the ionic well. First-principles calculation and Landau theory provide a basis for understanding these trends. The piezoelectricity is enhanced by a small anharmonicity of the ionic potential well, while SHG is enhanced by a large anharmonicity of the electronic potential well. Here, harmonicity refers to the potential energy of the electron being proportional to the square of the displacement of the electron from its equilibrium position within the well, whereas in this work, anharmonicity will refer specifically to the potential energy term proportional to the cubic power of such electron displacements. (Higher order terms in such a Taylor expansion of the electron potential energy versus its displacement are also considered anharmonic but they will not be considered here). This work highlights the design paradigm of achieving both enhanced nonlinear optical properties and piezoelectric response by engineering anharmonicity differently in different frequency regimes. Because the polarization in $Zn_{1-x}Mg_xO$ is switchable, precise domain control can be achieved in these thin films down to 800 nm domain size, which enables optical quasi-phase-matched (QPM) SHG and domain-engineered electro-optic and piezoelectric devices with a materials growth process that is CMOS-compatible.[34,35]

## 2. Results and Discussion

$Zn_{1-x}Mg_xO$ films ($x$ = 0, 16, 23, 28, 37 mol%) with a thickness of 150 nm were epitaxially grown at room temperature along the *c*-axis on (111)-Pt//(0001)-$Al_2O_3$ via RF magnetron sputtering from metal targets.[31] Over this range of $x$, $Zn_{1-x}Mg_xO$ adopts the wurtzite structure (**Figure 1a)** in PVD-grown thin films.[22,27,28] Mg substitutes on the Zn site due to their similar ionic radii and electronegativity, providing a local distortion of the bond lengths and angles.[31,36] **Figure 1b** shows the X-ray diffraction (XRD) and the film stack of (0001)-$Zn_{1-x}Mg_xO$//(111)-Pt//(0001)-$Al_2O_3$. $Zn_{1-x}Mg_xO$ maintains the wurtzite structure and high crystallinity without forming a secondary rocksalt phase.[22,31,36] The (0002)-$Zn_{1-x}Mg_xO$ peak shifts gradually toward a higher $2\theta$ in XRD with increasing Mg concentration. This indicates a systematic contraction in the *c*-lattice constant by Mg substitution, which agrees well with the previous studies.[22,27,31] The films are epitaxial to the underlying Pt electrode and have an out-of-plane full-width-half-max value in the omega x-ray circle between 1.5° and 2.0° at the lowest and highest Mg concentrations, respectively. An expansion along the *a* axis and an overall reduction in the *c*/*a* ratio have also been reported for comparable films.[22,27,31] Scanning electron microscopy (SEM) was performed to evaluate the surface microstructure, while surface roughness was measured by atomic force microscopy (AFM). Results reveal uniform faceted surface grains with an average diameter of ~ 100 nm and smooth surface (RMS roughness ~1.3 nm) (**Figure S1**, Supporting Information). These observations illustrate high-quality films for further optical and electromechanical studies. **Figure 1(c)** exhibits the hysteresis loop for $Zn_{0.63}Mg_{0.37}O$, demonstrating that the ferroelectricity emerges beyond the critical concentration (~34%).[31] Although the polarization in ZnO is not switchable, alloying with Mg has been proven to be an effective way to access the ferroelectricity without further external stimuli. The coercive field is found to be around 3 MV/cm, similar to that reported in previous studies.[31]

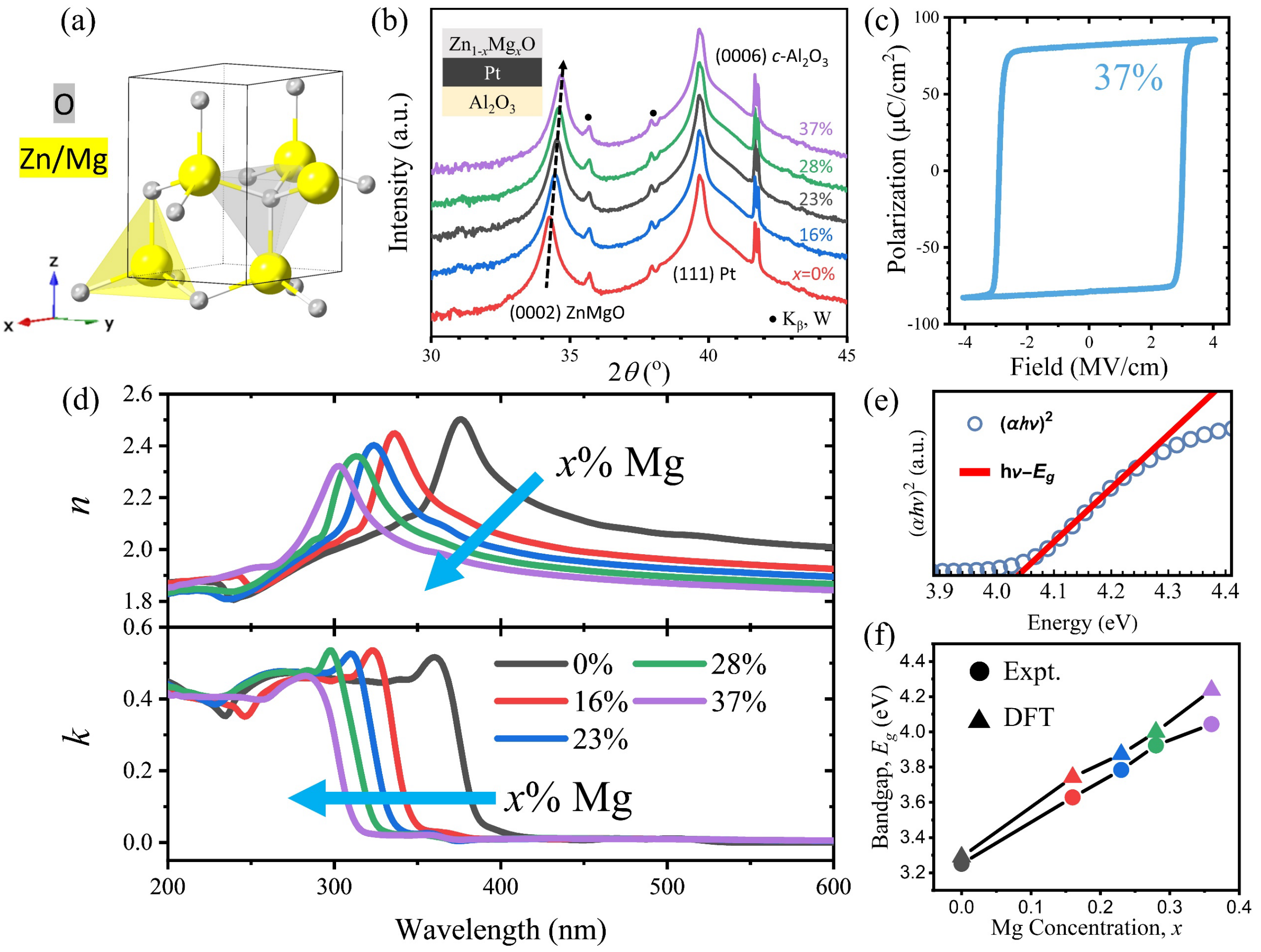

**Figure 1**. Crystal structure, XRD, and linear optical properties of $Zn_{1-x}Mg_xO$. (a) Wurtzite structure of $Zn_{1-x}Mg_xO$. (b) X-ray diffraction for a series Mg concentration ($x$ = 0, 0.16, 0.23, 0.28, and 0.37). The dashed line highlights the $Zn_{1-x}Mg_xO$ (0002) peak as a function of Mg concentration. Asterisks indicate $K_\beta$, W peaks. (c) P-E hysteresis loop for $Zn_{0.63}Mg_{0.37}O$. (d) Complex refractive indices $\tilde{n} = n + ik$ of $Zn_{1-x}Mg_xO$ from 200 nm to 600 nm. (e) Tauc fitting of $Zn_{0.63}Mg_{0.37}O$. (The dotted and solid lines represent experimental results and fit, respectively. (f) Bandgaps of $Zn_{1-x}Mg_xO$ as a function of Mg concentration. The circle represents experimental results, and the triangle stands for DFT.

The bandgap measurements are discussed first, then the SHG measurements, followed by the piezoelectric measurements. The complex linear optical refractive indices ($\tilde{n} = n + ik$) (**Figure 1d**) were studied using spectroscopic ellipsometry from 1800 nm to 200 nm (equivalent to 0.69 eV to 6.2 eV). Mg substitution reduces the refractive indices and pushes the band edge towards higher energy. Since the experimental bandgaps of ZnO and MgO are reported to be 3.2 eV and 7.8 eV,[37,38] respectively, Mg addition is expected to increase the bandgap. The refractive indices below the band edges were fitted using the birefringent Cauchy dispersion relation due to the uniaxial structure of $Zn_{1-x}Mg_xO$, and the Cauchy parameters are summarized in **Table S1** (Supporting Information). Using the Tauc method[39,40] and the measured $n$ and $k$ of $Zn_{1-x}Mg_xO$, a direct band transition was confirmed throughout the $Zn_{1-x}Mg_xO$ series from x = 0 to 0.37. As expected, adding Mg reduces the refractive indices and pushes the band edge towards higher energy. The Tauc fitting for $x$=0.37 is shown in **Figure 1e,** which results in the largest bandgap in the series. The extracted bandgaps as a function of Mg concentration from the experiment (**Figure 1f**) show a linear dependence on the Mg concentration, which agrees well with previous studies and is in agreement with the DFT predictions.[28,29]

Second harmonic generation (SHG) is a nonlinear optical process that converts two photons at ω frequency to one photon at 2ω frequency.[41] SHG has been widely applied in coherent lasing sources, structural characterization, and biological imaging. **Figure 2a** exhibits the experimental geometry of the SHG measurement. The incident polarization ($E^{\omega}$) is rotated by a half-wave plate (azimuthal angle $\varphi$), and both $p$- and $s$- polarized SHG intensities ($I_p^{2\omega}$ and $I_s^{2\omega}$) were collected as a function of incident polarization $\varphi$. The measured films are a stack of (0001)-$Zn_{1-x}Mg_xO$//(111)-Pt//(0001)-$Al_2O_3$.

Due to the strong reflection and absorption of $\omega$ and $2\omega$ frequencies in the Pt layer, multireflection and absorption of both the fundamental $\omega$ and second harmonic $2\omega$ waves (**Figure 2b**) need to be considered to correctly model and extract second-order nonlinear susceptibilities. It is noted that typical nonlinear optical analysis involves numerous assumptions, such as slowly varying amplitude approximation,[21] weak reflection of the nonlinear source wave,[42,43] and Kleinman's symmetry.[44–46] Excluding multiple reflections in the optical analysis would lead to a failure of available SHG models.[47,48] Kleinman's symmetry forces $d_{31}^{\mathrm{SHG}} = d_{15}^{\mathrm{SHG}}$ in Voigt notation in the $Zn_{1-x}Mg_xO$ system which could be problematic and therefore is not assumed *a priori* in the analysis presented in this current study.[41,49] Failure to take all necessary effects into account can result in one to two orders of magnitude error in the estimation of the SHG coefficients as compared with bulk single crystals. An advanced modeling tool named ♯SHAARP was employed, which fully accounts for multiple reflections, interference, and the complete anisotropic SHG tensor.[49]

The polarized second harmonic intensities (dots) and fitting (solid lines) are shown in **Figure 2c**, where blue and red represent *p*- and *s*- polarized intensity, respectively. The power-dependent SHG response shows a quadratic dependence between pump power and SHG intensity, $I^{2\omega} \propto (I^{\omega})^2$, confirming intrinsic SHG response from the $Zn_{1-x}Mg_xO$ films (**Figure S2**, Supporting Information). The absolute second harmonic susceptibilities of $Zn_{1-x}Mg_xO$ are extracted by referencing against a well-studied $LiNbO_3$ single crystal and are summarized in **Figure 2d**, **Table S2, and Equation S7-25** (Supporting Information). A 0.5mm thick ZnO single crystal was also studied to verify and benchmark the analysis. The absolute SHG susceptibility $d_{33}^{\mathrm{SHG}}$ of the ZnO single crystal is found to be 7.30 pm $V^{-1}$, which agrees well with reported values in the literature.[45,46,50] This indicates a robust and reliable methodology using #SHAARP[49] for the characterization of nonlinear optical response (**Figure S3**, Supporting Information). **Figure 2d** summarizes the absolute SHG susceptibilities of $Zn_{1-x}Mg_xO$ as a function of Mg concentration, where $E_g$ increases monotonically with the Mg concentration. Interestingly, a nearly 50% enhancement of $d_{33}^{\mathrm{SHG}}$ is found in $x$ = 0.16 and 0.23 as compared to pure ZnO films and single crystals. This enhancement is significantly less than the enhancement of 420% reported in literature;[25] Such large previously reported SHG coefficients (approaching ~50pm/V) for wide bandgap semiconductors such as ZnO are unusual. A slightly different wavelength or the details of the film growth are likely minor contributors to this discrepancy since the pure ZnO film in our current study exhibits SHG coefficients similar to the bulk crystal. The most likely source of the discrepancy we surmise is the modeling of the SHG response in the thin film. For example, if we assume Kleiman's symmetry in our case, it can erroneously result in ~50 pm/V of $d_{33}^{\mathrm{SHG}}$ for both $x$=0.28 and 0.36, suggesting a possible reason for such discrepancy with the previous literature (**Figure S6**, Supporting Information). Given our confidence in taking care of all the relevant details in modeling using the proven #SHAARP code and benchmark analysis using both films and the single crystal, the prior reported enhancement needs to be revisited.[25] Further Mg substitution beyond 23% Mg tends to suppress $d_{33}^{\mathrm{SHG}}$ before reaching the maximum solubility. On the other hand, both the raw $d_{31}^{\mathrm{SHG}}$ and $d_{15}^{\mathrm{SHG}}$ exhibit a monotonic reduction with increasing Mg concentration; nonetheless as shown below, the electronic anharmonicity increases with the Mg concentration for these coefficients as well.

In the classical nonlinear spring model, the energy dispersion of the SHG nonlinear susceptibility is given by, $d_{ij}^{\mathrm{SHG}} \propto A_{\mathrm{e},ij}\left(E_g^2 - 4{E_{\mathrm{photon}}}^2\right)^{-1}\left(E_g^2 - {E_{\mathrm{photon}}}^2\right)^{-2}$, where $d^{\mathrm{SHG}}$, $A_{\mathrm{e},ij}$, $E_g$ and $E_{\mathrm{photon}}$ are the SHG susceptibility, anharmonicity of the nonlinear spring, bandgap, and probing photon energy.[21,41] In particular, $A_{\mathrm{e},ij}$ is the strength of the lowest power anharmonic term in the energy profile for a bound electron oscillating with respect to its nuclei. The anharmonicity, $A_{\mathrm{e},ij}$, describes the intrinsic origin of SHG response and captures the physics underlying design methods for highly efficient NLO crystals, including triangle-planar anion groups, second-order Jahn-Teller effects, and lone electron pair cations.[15,51] Based on this dependence, $d^{\mathrm{SHG}}$ tends to be suppressed by the larger bandgap at higher Mg concentrations.[52] However, nearly 50% enhancement of $d_{33}^{\mathrm{SHG}}$ up to 23% Mg indicates a significant enhancement in the SHG coefficients, which is unexpected. By correcting the raw $d_{ij}^{\mathrm{SHG}}$ data in **Figure 2d** for the dispersion term, $\left(E_g^2 - 4E_{\mathrm{photon}}{}^2\right)^{-1}\left(E_g^2 - E_{\mathrm{photon}}{}^2\right)^{-2}$, one can obtain a quantity proportional to the anharmonicity of the electronic well, $A_{\mathrm{e},ij}$ as shown in **Figure 2e**. The increase of $A_{\mathrm{e},ij}$ for all three susceptibilities suggests that Mg substitution promotes the anharmonicity of the electronic potential well along both the ordinary and the extraordinary polarization directions. In particular, a roughly 400% enhancement of the $A_{\mathrm{e},33}$ is observed, indicating a more significant influence on the electronic potential well along the polar direction. Simultaneously, the spontaneous polarization, $P_s$ decreases as a function of Mg concentration experimentally (**Figure 2f**);[31] DFT calculations attribute this trend mainly to the reduced Born effective charge of Zn and Mg, from 2.17 in ZnO to an average of 2.03 in $Zn_{0.61}Mg_{0.39}O$ along the *c* direction. Thus, the substantial enhancement of $d_{33}^{\mathrm{SHG}}$ and $A_{\mathrm{e},33}$ in this material system are both found to be inversely correlated to the change in the spontaneous polarization, contrary to conventional expectations,[53–55] suggesting that it is the electronic anharmonicity rather that the spontaneous polarization that determines the SHG coefficients.

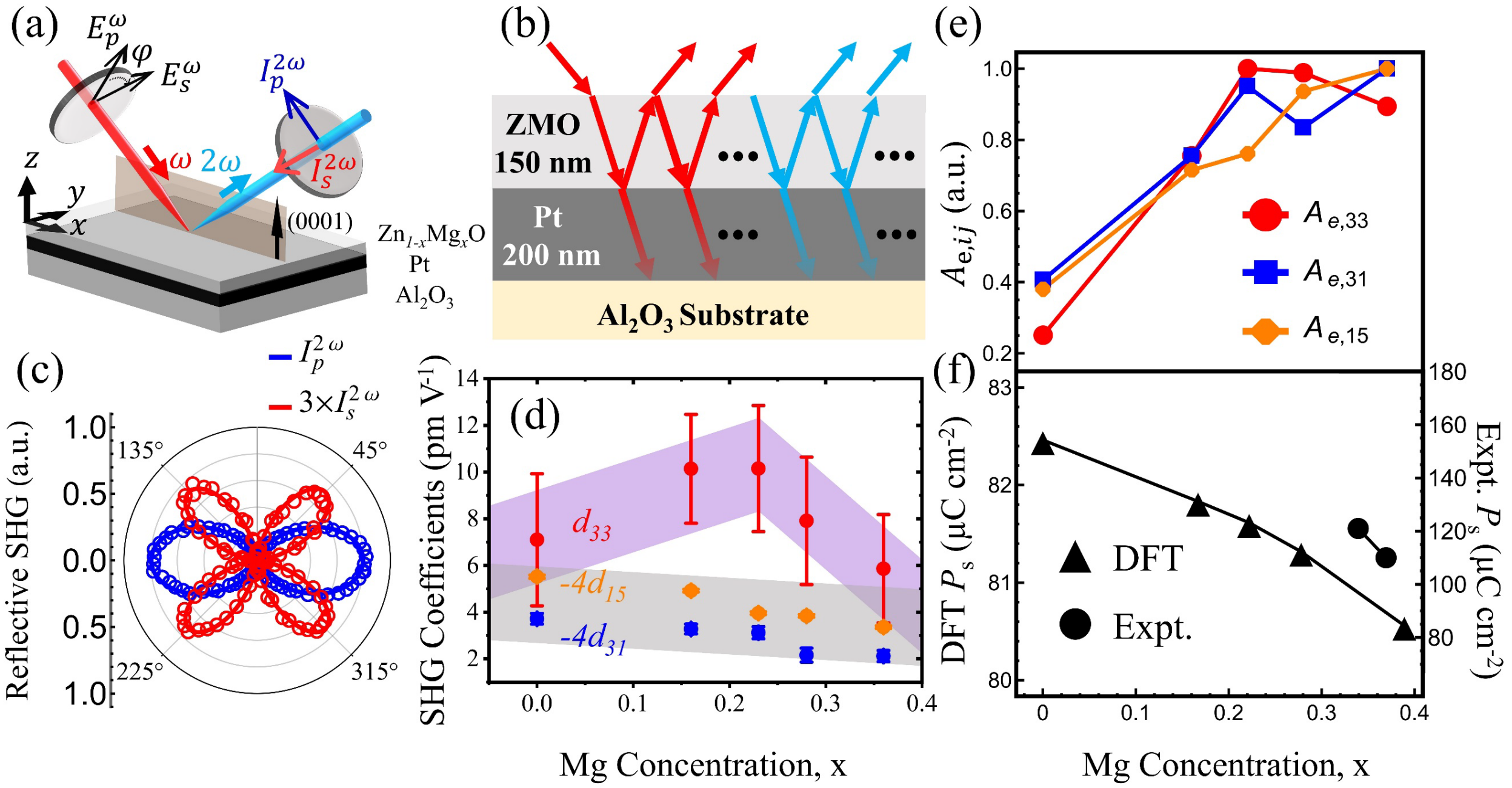

**Figure 2**. Nonlinear optical properties of $Zn_{1-x}Mg_xO$. (a) Experimental geometry of SHG. Red and blue rays are the fundamental wavelength and SHG wavelength, respectively. (b) Multireflection model for SHG analysis. Red and blue waves correspond to the fundamental and SHG waves. ZMO refers to $Zn_{1-x}Mg_xO$. (c) SHG polarimetry of $Zn_{1-x}Mg_xO$. Blue and red correspond to *p*- and *s*- polarized second harmonic intensity. The dotted and solid lines represent the experimental results and fit separately. (d) Absolute SHG coefficients of $Zn_{1-x}Mg_xO$ films. $d_{33}^{\mathrm{SHG}}$, $d_{31}^{\mathrm{SHG}}$ and $d_{15}^{\mathrm{SHG}}$ are colored in red, blue, and orange. The background trendlines are a guide to the eye. (e) Relative anisotropic anharmonicity $A_{\mathrm{e},ij}$ change for three SHG coefficients using the same color scale as (d). (f) Spontaneous polarization from DFT and remanent polarization from the experiment. Triangles and circles represent DFT and experimental results.

**Figure 3a** summarizes the $d^{\mathrm{SHG}}$ and $E_g$ among state-of-art nonlinear optical (NLO) materials.[20] With increasing bandgaps, the magnitudes of nonlinear optical susceptibilities reduce dramatically. Remarkably, $Zn_{1-x}Mg_xO$ with *x* from 0% to 23% clearly demonstrates a substantial enhancement of both $d^{\mathrm{SHG}}$ and $E_g$, contrary to the overall general trend (highlighted in grey) observed across a broad range of material families.

To develop some intuitive understanding of the mechanism that supports this contrary trend, let us consider a classical anharmonic electronic potential well given by $U_e = \frac{mE_g^2}{2\hbar^2}x^2 + \frac{mA_e}{3}x^3 + \cdots$, where $U_e$, $E_g$, $A_e$, $m$, $\hbar$, and $x$ are, respectively, the electronic potential energy, electronic bandgap, anharmonicity of the oscillator, effective mass, reduced Planck constant, and the relative position between the electron and nuclei.[21,41] Using the classical theory of anharmonicity, one can derive expressions for the dielectric susceptibility ($\chi_e$) and $d^{\mathrm{SHG}}$ as a function of $E_g$ and $A_e$ for the non-resonant SHG process as shown in **Equation (1)** and **(2)**: [21,41]

$$\chi_e = \frac{N_\mathrm{e} e^2 \hbar^2}{\varepsilon_0 m E_g^2 - \varepsilon_0 m E_{\mathrm{photon}}^2} \tag{1}$$

$$d^{\mathrm{SHG}} = \frac{N_\mathrm{e} A_e e^3 \hbar^6}{m^2 (E_g^2 - 4E_{\mathrm{photon}}{}^2)(E_g^2 - E_{\mathrm{photon}}{}^2)^2} \tag{2}$$

where $\varepsilon_0$ is the vacuum permittivity, $N_\mathrm{e}$ is the number of dipoles per unit volume, $e$ is the electron charge, and $E_{\mathrm{photon}}$ is the experimental pumping energy at 0.8 eV. Experimental observation has confirmed a ~16% increase in the $E_g$ and 400% enhancement in the $A_{e,33}$ from $x$ = 0% to 23%. **Figure 3b** and **c** illustrates the relative changes of $\chi_e$ and $d^{\mathrm{SHG}}$ as compared with pure ZnO following **Equation 1** and **2**. With increasing $E_g$, the $\chi_e$ decreases due to the reduction of the electronic polarizability, as seen in **Figure 3b**. Normally, an increasing $E_g$ tends to suppress the magnitude of $d^{\mathrm{SHG}}$ for a given $A_e$ as seen in **Figure 3a**. However, in $Zn_{1-x}Mg_xO$, a substantial enhancement of $A_{e,33}$ offsets the effect of increasing $E_g$ and decreasing $\chi_e$ to yield a net increase in $d^{\mathrm{SHG}}$, as indicated in **Figure 3c**. Increasing Mg concentration tends to reduce the Zn-O bond length, as confirmed by the lattice parameters.[31] Since the decrease in the bond length will promote the Coulomb repulsion between Zn(Mg) and O, it is postulated that the distorted O-2$p_z$ and the Zn 3*d* and Mg 2*p* orbitals are likely the driving force for the enhanced nonlinearity in the $Zn_{1-x}Mg_xO$ system. These orbitals dominate the density of states near the band edge, and thus are the major source for the non-resonant SHG response near it.[56,57]

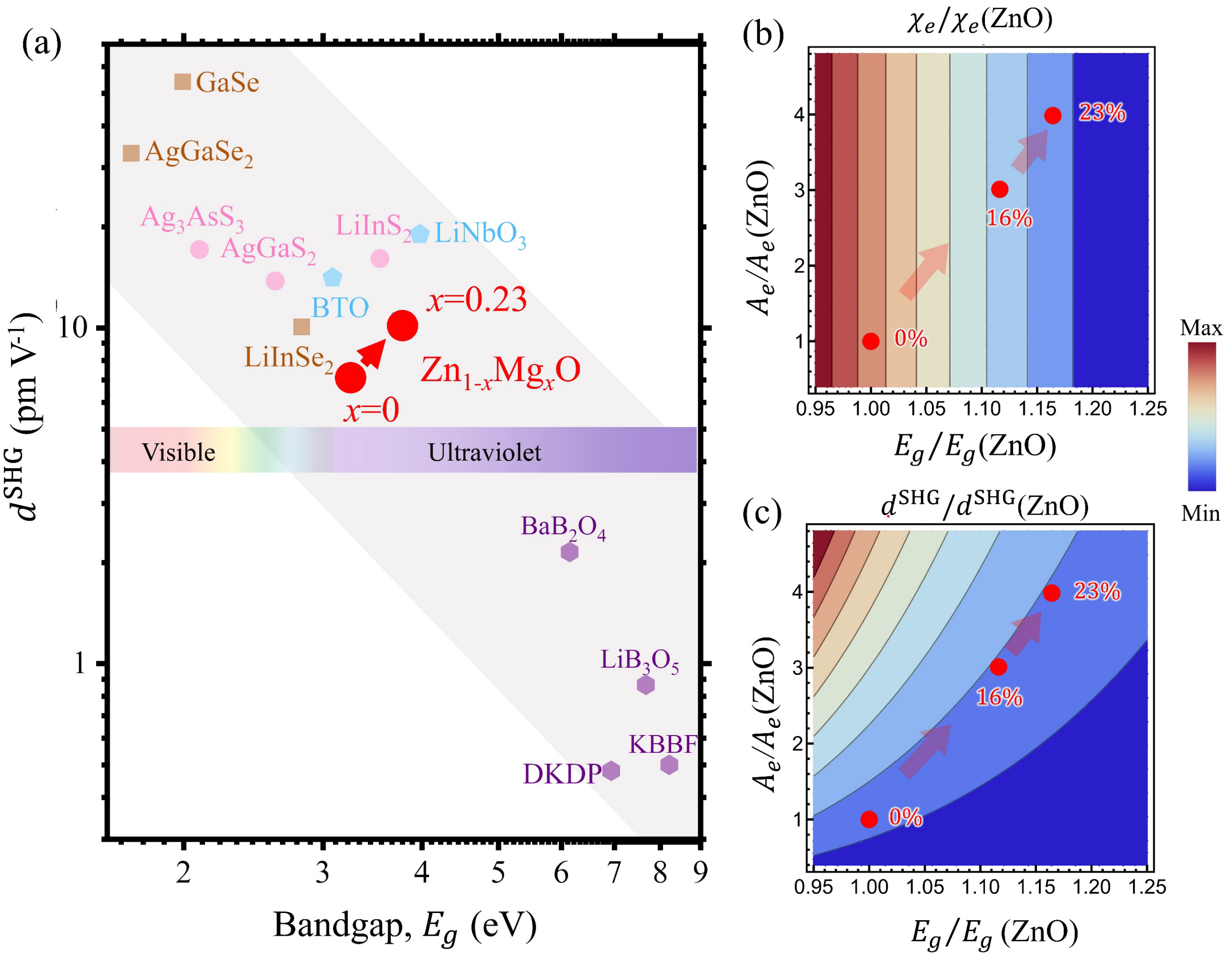

**Figure 3**. Competing effects among bandgap ($E_g$), anharmonicity ($A_e$), and SHG susceptibility ($d^{\mathrm{SHG}}$). (a) Summary of SHG coefficients and bandgaps among various NLO materials and $Zn_{1-x}Mg_xO$. The labels BTO, DKDP, and KBBF stand for $BaTiO_3$, $KD_2PO_4$, and $KBe_2BO_3F_2$, respectively. The grey region is a guide to the eyes of the relationship between $d^{\mathrm{SHG}}$ and $E_g$. (b-c)

Contour plot of (b) electronic dielectric susceptibility $\chi_e$ and (c) SHG coefficients $d^{\mathrm{SHG}}$ as a function of normalized $E_g$ and $A_{e,33}$ relative to the ZnO using **Equation 1** and **2**. $N_e$, $m$, $\hbar$, $e$ and $\varepsilon_0$ are set to unit values. The red arrows show the trajectory of the change from $x$ = 0 to 23% in $Zn_{1-x}Mg_xO$.

Next, the enhancement in the piezoelectric coefficients is discussed. **Figure 4** summarizes the piezoelectric response and writing of periodically poled ferroelectric domains in $Zn_{1-x}Mg_xO$ using a combination of interferometric displacement sensing (IDS) and band excitation piezoelectric force microscopy (BE-PFM). The as-grown films exhibit uniform PFM phase across the scanned area, indicating that the films were single domain as-deposited with a spontaneous polarization pointing down in their initial state (**Figure S4**, Supporting Information). The piezoelectric coefficient ( $d_{33}^{\mathrm{Piezo}}$ ), extracted from the out-of-plane electromechanical displacements measured via IDS as a function of applied AC voltage, are shown in **Figure 4a** across the $Zn_{1-x}Mg_xO$ concentration series. Here, nearly a two-fold increase in $d_{33}^{\mathrm{Piezo}}$ is observed between Mg concentrations of $x$=0.0 and $x$=0.37, consistent with the enhancement observed in prior literature.[22–24] The discrepancy in the magnitude of $d_{33}^{\mathrm{Piezo}}$ between our work and other studies is likely due to the clamping effect from the substrate in our thinner epitaxial films.

To understand this enhancement in $d_{33}^{\mathrm{Piezo}}$, a Landau-Ginzburg-Devonshire (LGD) theory prediction of the piezoelectric coefficient is utilized.[58–60] Starting from the free energy of a system, one can derive the piezoelectric coefficients, given as $d^{\mathrm{Piezo}} = 2QP_s\chi_i = \frac{Q}{\varepsilon_0 A_i}$, where $\chi_i$ is the linear ionic susceptibility tensor, $Q$ is the electrostrictive tensor and $A_i$ is the anharmonicity tensor of the ionic potential well of the Zn or Mg ions in the oxygen tetrahedral cage (**Equation**

**S1-6 and Figure S5**, Supporting Information). According to the Gibbs free energy of ferroelectrics, $A_i$ represents the magnitude of cubic term in the Taylor expansion of the ionic potential energy versus ionic displacement away from its equilibrium ionic position at $P = P_s$. Based on the changes in $d^{\mathrm{Piezo}}$, $P_s$ and $\chi_i$, the ionic anharmonicity and electrostrictive coefficient can be quantitatively analyzed.[31] **Figure 4b** highlights the changes in $Q$, $A_i$ and the corresponding $d^{\mathrm{Piezo}}$ among $Zn_{1-x}Mg_xO$ concentrations. Between $x$=0 and $x$= 37%, $Q$ varies only slightly (less than 5%), indicating that the electrostrictive response is not the major factor enhancing the piezoelectric response. On the other hand, $P_s$ from DFT exhibits a ~2% reduction (**Figure 2f**) and the reported ionic dielectric susceptibility, $\chi_i$, exhibits a ~75% enhancement,[31] resulting in ~40% reduction in the anharmonicity, $A_i$ of the ionic potential well between $x$ = 0 and 0.37. ($A_i = \frac{1}{2\varepsilon_0 P_S \chi_i}$ as derived in the supplementary section 7, Supporting Information) The reduced ionic anharmonicity combined with a slight increase in the electrostriction produces a near 200% enhancement in the piezoelectric coefficient. The smaller anharmonicity $A_i$ of the ionic potential well has a similar effect in $Zn_{1-x}Mg_xO$ as the double well potential flattening on approaching a phase boundary in perovskite piezoelectrics (**Figure S5**, Supporting Information).[58,62–65] Applying LGD theory in $Zn_{1-x}Mg_xO$ also predicts the lowering of the energy barrier of ferroelectric switching with increasing Mg concentration (**Figure S5c**, Supporting Information). This further provides insight into the polarization switching mechanism from non-ferroelectric ZnO towards ferroelectric $Zn_{1-x}Mg_xO$. That is, the softening of the wurtzite structure (smaller *c*/*a* ratio) and flattened double well potential lowers the energy barrier of the polarization reversal, facilitating cation motion through the oxygen at the base of the tetrahedron to reverse the spontaneous polarization. This theoretical framework agrees well with experimental observations of enhanced dielectric constant,[31] increased piezoelectric response, reduction in spontaneous polarization, and switchable ferroelectric polarization with increasing Mg concentration in $Zn_{1-x}Mg_xO$, as demonstrated in this study.

To highlight the potential technological interest in this material, piezoelectric force microscopy was utilized to demonstrate fine control over domain patterning in ferroelectric $Zn_{1-x}Mg_xO$. Domain reversal with electric fields is far simpler and preferred as compared with domain reversal during synthesis using engineered surface termination in AlN,[66,67] and polarity control in ZnO.[68] Coherence length ($l_c$) is the size of the domain in a domain grating of period $2l_c$ required for quasi-phase-matched SHG, defined as $l_c = \lambda^\omega/4(n^{2\omega} - n^\omega)$, where $\lambda$ and $n$ are wavelength and refractive index, and superscript ω represents the corresponding frequency.[69] The calculated $l_c$ as a function of fundamental wavelength $\lambda^\omega$ in $Zn_{0.63}Mg_{0.37}O$ is illustrated in **Figure 4c**. The $l_c$ is found to be 800 nm when a wavelength of $\lambda^\omega$ = 650 nm is halved to $\lambda^{2\omega}$ = 325 nm. The theoretical limit of $l_c$ is found to be ~650 nm when $\lambda^{2\omega}$ approaches the band edge, the limit of the useful range for nonlinear optics. As proof of the feasibility of QPM in $Zn_{1-x}Mg_xO$, a periodic poled pattern with a domain width of 800 nm is demonstrated in $Zn_{0.63}Mg_{0.37}O$. The BE-PFM amplitude (**Figure 4d**) and phase (**Figure 4e**) show a clear domain contrast with opposite polarization, demonstrating the realization of precise and controllable QPM for any targeted wavelength below the bandgap.

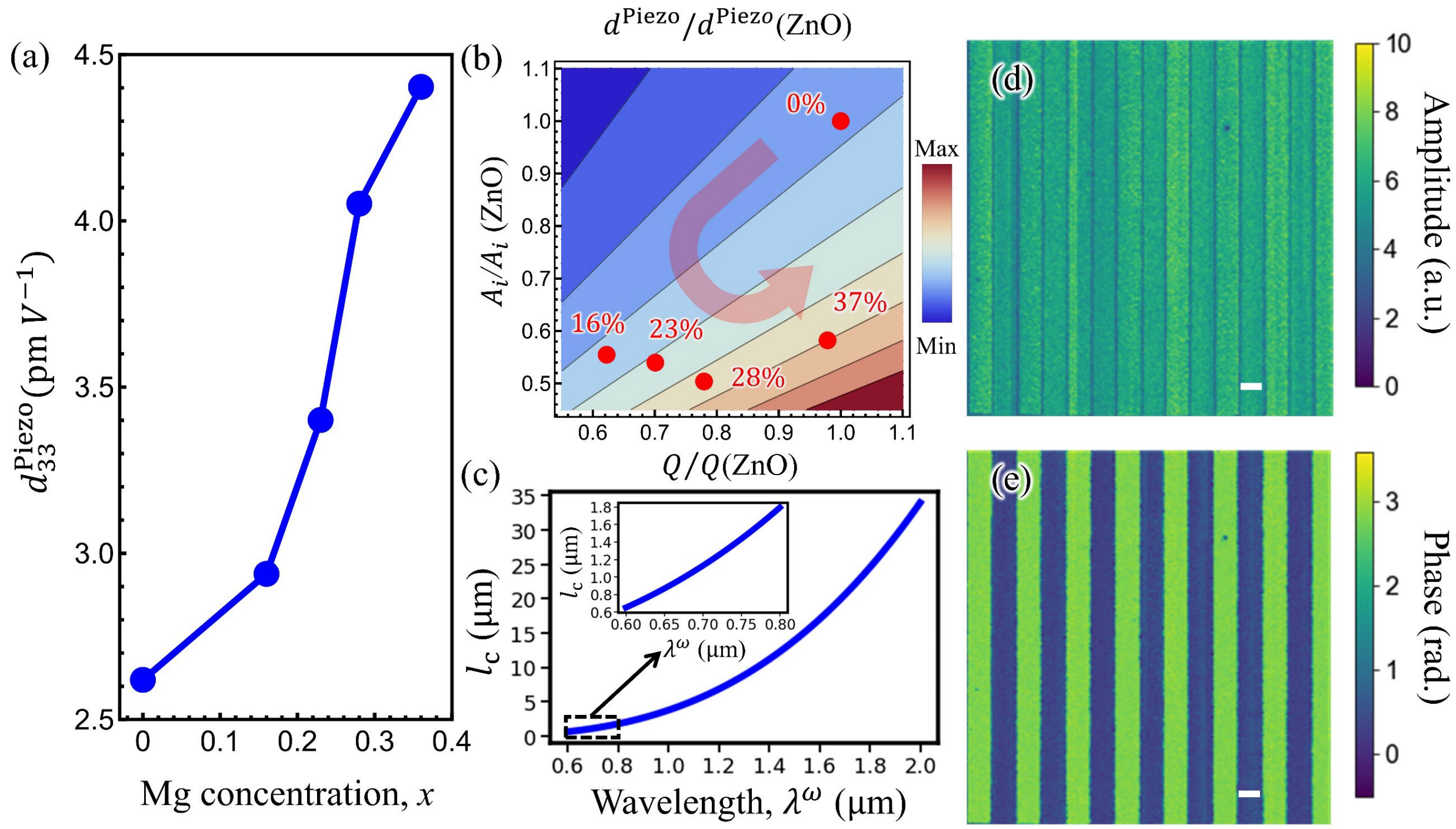

**Figure 4**. PFM results and QPM of $Zn_{1-x}Mg_xO$. (a) $d_{33}^{\mathrm{Piezo}}$ acquired from interferometric displacement sensing PFM on $Zn_{1-x}Mg_xO$ series. (b) Contour plot of $d^{\mathrm{Piezo}}$ as a function of $Q$ and $A_{\mathrm{i}}$ relative to the ZnO. $\varepsilon_0$ is set to the unit value. The red arrow indicates the changes from $x$=0 to 0.37 in $Zn_{1-x}Mg_xO$. (c) Coherence length ($l_c$) as a function of the fundamental wavelength $\lambda^{\omega}$ of $Zn_{0.63}Mg_{0.37}O$ for QPM SHG devices. The inset shows a zoomed-in wavelength range between 0.6 – 0.8 μm. (d-e) BE-PFM mapping of periodically poled $Zn_{0.63}Mg_{0.37}O$ with a scan size of 12.8 × 12.8 μm poled at ±50V. (d) is the amplitude, and (e) is the phase. The scale bars in panels (d) and (e) are 0.8 μm.

## 3. Conclusions

In summary, $Zn_{1-x}Mg_xO$ brings exceptional optical, electrical, and electromechanical responses to life in a single multifunctional platform. The simultaneous increase in both the second-order nonlinear optical susceptibility and the optical bandgap in $Zn_{1-x}Mg_xO$ suggests that increasing the anharmonicity of the electron potential well can offset a bandgap increase for designing NLO materials in the ultraviolet spectral region. On the low-frequency end, adding Mg reduces the anharmonicity of the ionic potential well, leading to an enhancement of the piezoelectric response. This study thus demonstrates that the anharmonicity can be independently engineered in the optical and low-frequency regimes through chemical pressure. This may be key to optimizing exceptional piezoelectric and nonlinear optical response in one material. Moreover, precisely controlled periodic ferroelectric domain patterns for optical quasi-phase-matching (QPM) were demonstrated in $Zn_{1-x}Mg_xO$, which opens new possibilities for NLO optical waveguides for efficient nonlinear optical conversion and all-optical switching schemes deeper into the ultraviolet range. The presence of ferroelectricity with large polarization, high ionic dielectric susceptibility, improved piezoelectric response, large tunable bandgap, and significantly enhanced SHG response make $Zn_{1-x}Mg_xO$ a promising candidate for applications such as microelectromechanical systems, integrated optics, nonlinear photonics, and strain-tunable photonics.

## 4. Experimental Section

*Growth of $Zn_{1-x}Mg_xO$ films, XRD, SEM, EDS, AFM*: $Zn_{1-x}Mg_xO$ thin films were grown via a radio frequency (RF) magnetron co-sputtering technique using metallic Zn and Mg targets. A 2-inch (0001)-$Al_2O_3$ wafer was rinsed successively in isopropanol, methanol and acetone for 1 min, followed by an ultraviolet ozone treatment for 10 min. The wafer was placed in a sputtering chamber until the base pressure was reduced to $1 \times 10^{-7}$ Torr at 300°C. Deposition of a 200 nm thick (111)-Pt layer on (0001)-$Al_2O_3$ was performed at 10 mTorr in Ar atmosphere. $Zn_{1-x}Mg_xO$ thin films were then sputtered on the (111)Pt//(0001)-$Al_2O_3$ substrate at room temperature. During depositions, a gas flow of Ar and O was fixed to 16 sccm and 4 sccm respectively, and a total pressure was kept to 4 mTorr. The Mg content was controlled by changing the Mg target power between 0 to 48 W, while the Zn target power was fixed at 23 W. The chemical composition and surface morphology were confirmed by energy dispersive spectroscopy (EDS) and scanning electron microscope (SEM) in Zeiss Sigma. Crystal structure was investigated via x-ray diffraction (XRD) using Cu–Kα1 (1.5406 Å) in a Panalytical Empyrean, and the surface roughness was measured using atomic force microscope (AFM) in Asylum MFP3D.

*Optical Spectroscopic Ellipsometry*: Optical spectroscopic ellipsometry was conducted using Woollam M-2000 and Woollam M-2000F spectroscopic ellipsometers. Woollam M-2000F has a fixed incident angle of around 60°, and three incident angles (55°, 65°, and 75°) were collected using Woollam M-2000. The two results are cross confirmed and combined to reveal the complex refractive indices from 200 nm – 1800 nm using the same model.

*Tauc method for optical bandgap*: The Tauc equation is expressed as $(\alpha h\nu)^{1/n} = A(h\nu - E_g)$, where $\alpha$, $h$, $\nu$, $A$ and $E_g$ are the absorption coefficient, Planck's constant, photon frequency, proportionality constant, and bandgap.[40] $n$ is the measure of direct or indirect transition and $n$ is set to 0.5 for the best fitting condition in this study.

*SHG Measurement*: SHG has been widely used to confirm noncentrosymmetric structures, ferroelectric response, and wavelength conversion efficiency. The polarization-resolved SHG measurements were carried out at room temperature in 45-degree reflection mode on samples. The SHG measurement is an all-optical technique where two photons of frequency ω with fields $E_j$ and $E_k$ and polarization directions $j$ and $k$, respectively, interact with a material with a non-zero $d_{ijk}$ tensor and generate a polarization $P_i^{2\omega}$ of frequency $2\omega$ in the $i$ direction. The SHG intensity, $I^{2\omega}$, was detected with a Hamamatsu photomultiplier tube. A Ti-sapphire laser (Spectra-Physics) with an output of 800 nm, 80 fs pulses at 2 kHz frequency was used. The fundamental light with a central wavelength at 1550 nm is generated through an optical parametric amplifier after the Coherent Libra Amplified Ti: Sapphire femtosecond laser system (85 fs, 2 kHz). Here, $(x, y, z)$ is the lab coordinate system where experiments are performed. The plane of incidence is defined in the $x$-$z$ plane, and $z$ corresponds to the surface normal. The surface normal of $Zn_{1-x}Mg_xO$ is the (0001) plane.

*Electronic-Structure Calculations*: First-principles calculations were performed using norm-conserving pseudopotentials within the Perdew–Burke–Ernzerhof (PBE) approximation to the exchange-correlation energy.[70–78] Supercell structures were generated using the Atomic Simulation Environment (ASE) module.[79] To consider atomic fraction of $x$ = 0-40%, a supercell size of 3 × 3 × 2 was considered with minimal Mg clustering. The $\boldsymbol{k}$-point spacing in the first Brillouin zone was set to 0.05 $Å^{-1}$, and the kinetic energy cutoff for the electronic wavefunctions was set to 80 Ry with a charge density cutoff of 320 Ry. The lattice parameters and $c/a$ ratios of the unit cell of ZnO and the supercell structures of $Zn_{1-x}Mg_xO$ were calculated using geometry optimization. The total energy and force thresholds were set to $10^{-5}$ Ry and $10^{-4}$ Ry/bohr, respectively. Band gaps were calculated within the DFT+$U$ approximation at fixed (PBE) geometry.[80] Unlike the empirical approach that consists of adjusting the Hubbard $U$ parameters to match experimental bandgaps, the Hubbard $U$ parameters were calculated using an entirely nonempirical method *via* density functional perturbation theory (DFPT).[81–83] A detailed description of this nonempirical DFT+$U$ framework is provided in the supplementary information. Following Ref.[82], the Hubbard $U$ parameters were applied to O-2$p$ orbitals (cf. Tables S3 and S4). Using this approach, the spontaneous polarization was calculated *via* the Berry phase method.[84,85]

*IDS-PFM, BEPFM measurement, periodic poling*: Interferometric displacement sensing PFM measurements were taken using an Oxford Instruments Cypher atomic force microscope equipped with a Polytec OFV-5000 Modular Vibrometer routed to the tip for measuring tip displacements, i.e., the piezoelectric coefficient ($d_{33}^{\mathrm{Piezo}}$). AC voltages ranging from 2 to 10 V at 40 kHz were used for a linear extraction of $d_{33}^{\mathrm{Piezo}}$, i.e. the slope of IDS-PFM amplitude versus applied AC voltage. The band excitation PFM measurements were acquired using an Oxford Instruments Cypher atomic force microscope with an imaging AC voltage of 2V, and periodic poling voltage of ±50V. Details describing band excitation functionality can be found elsewhere.[86] For all atomic force microscopy measurements, Budget Sensor Electri Muli75-G Cr/Pt coated probes were used.

*Remanent Polarization Measurement*: Remanent polarization values for Pt/ZMO/Pt capacitors were extracted from bipolar Polarization-Electric field (P-E) hysteresis loops driven at ≥ 4 MV/cm with a 100 Hz triangle wave at room temperature. A precision Multiferroic II tester (Radiant Technologies) was used to measure the P-E loops and to extract the remanent polarization values. Each reported value is averaged over 10 measurements on 5 capacitors.

## Supporting Information

Supporting Information is available from the Wiley Online Library or from the authors.

## Acknowledgments

R.Z., G.R., S.B., L.C.J., I.D., S.T-M., J-P. M. and V. G. were primarily supported as part of the center for 3D Ferroelectric Microelectronics (3DFeM), an Energy Frontier Research Center funded by the U.S. Department of Energy (DOE), Office of Science, Basic Energy Sciences under Award Number DE-SC002111, for the new materials development. R.Z. also received support from the NSF MRSEC Center for Nanoscale Science, DMR-2011839, for optical characterizations. The Piezoresponse Force Microscopy research was supported by the Center for Nanophase Materials Sciences (CNMS), which is a US Department of Energy, Office of Science User Facility at Oak

Ridge National Laboratory. B.W. and L.-Q.C. are supported by the National Science Foundation under the grant number DMR-2133373. Computational resources were provided through the Pennsylvania State University's Roar supercomputers within the Institute for Computational and Data Sciences. J.H. acknowledges support from NSF DMR-2210933.

## Conflict of Interest

The authors declare no conflict of interest.

## Data Availability Statement

The data that support the findings of this study are available from the corresponding author upon reasonable request.

## Reference

[1] N. L. Bishop, V. Kradinov, P. B. Reid, T. N. Jackson, C. T. DeRoo, S. Trolier-McKinstry, *J. Astron. Telesc. Instrum. Syst.* **2022**, *8*, 029004.
[2] R. H. T. Wilke, R. L. Johnson-Wilke, V. Cotroneo, S. McMuldroch, P. B. Reid, D. A. Schwartz, S. Trolier-McKinstry, *IEEE Trans. Ultrason. Ferroelectr. Freq. Control* **2014**, *61*, 1386.
[3] L.-H. Peng, C.-W. Chuang, L.-H. Lou, *Appl. Phys. Lett.* **1999**, *74*, 795.
[4] D. F. Nelson, E. H. Turner, *J. Appl. Phys.* **1968**, *39*, 3337.
[5] S. Kim, V. Gopalan, *Appl. Phys. Lett.* **2001**, *78*, 3015.
[6] N. Malkova, V. Gopalan, *Phys. Rev. B* **2003**, *68*, 245115.
[7] N. Malkova, S. Kim, V. Gopalan, *Appl. Phys. Lett.* **2003**, *83*, 1509.
[8] D. Scrymgeour, N. Malkova, S. Kim, V. Gopalan, *Appl. Phys. Lett.* **2003**, *82*, 3176.
[9] S. Trolier-McKinstry, S. Zhang, A. J. Bell, X. Tan, *Annu. Rev. Mater. Res.* **2018**, *48*, 191.
[10] X. Liu, P. Tan, X. Ma, D. Wang, X. Jin, Y. Liu, B. Xu, L. Qiao, C. Qiu, B. Wang, W. Zhao, C. Wei, K. Song, H. Guo, X. Li, S. Li, X. Wei, L.-Q. Chen, Z. Xu, F. Li, H. Tian, S. Zhang, *Science* **2022**, *376*, 371.
[11] F. Li, M. J. Cabral, B. Xu, Z. Cheng, E. C. Dickey, J. M. LeBeau, J. Wang, J. Luo, S. Taylor, W. Hackenberger, L. Bellaiche, Z. Xu, L.-Q. Chen, T. R. Shrout, S. Zhang, *Science* **2019**, *364*, 264.
[12] F. Li, D. Lin, Z. Chen, Z. Cheng, J. Wang, C. Li, Z. Xu, Q. Huang, X. Liao, L.-Q. Chen, T. R. Shrout, S. Zhang, *Nat. Mater.* **2018**, *17*, 349.
[13] T. Handa, R. Hashimoto, G. Yumoto, T. Nakamura, A. Wakamiya, Y. Kanemitsu, *Sci. Adv.* **2022**, *8*, eabo1621.
[14] T. W. Kasel, Z. Deng, A. M. Mroz, C. H. Hendon, K. T. Butler, P. Canepa, *Chem. Sci.* **2019**, *10*, 8187.

[15] H. Wu, H. Yu, Z. Yang, X. Hou, X. Su, S. Pan, K. R. Poeppelmeier, J. M. Rondinelli, *J. Am. Chem. Soc.* **2013**, *135*, 4215.
[16] D. K. T. Chu, H. Hsiung, *Appl. Phys. Lett.* **1992**, *61*, 1766.
[17] C. Liu, K. Gao, Z. Cui, L. Gao, D.-W. Fu, H.-L. Cai, X. S. Wu, *J. Phys. Chem. Lett.* **2016**, *7*, 1756.
[18] R. C. Miller, *Phys. Rev.* **1964**, *134*, A1313.
[19] B. Sahraoui, R. Czaplicki, A. Klöpperpieper, A. S. Andrushchak, A. V. Kityk, *J. Appl. Phys.* **2010**, *107*, 113526.
[20] D. N. Nikogosyan, *Nonlinear Optical Crystals: A Complete Survey*, Springer Science & Business Media, **2006**.
[21] R. W. Boyd, D. Prato, *Nonlinear Optics*, Academic Press, Amsterdam ; Boston, **2008**.
[22] X. Kang, S. Shetty, L. Garten, J. F. Ihlefeld, S. Trolier-McKinstry, J.-P. Maria, *Appl. Phys. Lett.* **2017**, *110*, 042903.
[23] Y. J. Chen, S. Brahma, C. P. Liu, J.-L. Huang, *J. Alloys Compd.* **2017**, *728*, 1248.
[24] S. Goel, B. Kumar, *J. Alloys Compd.* **2020**, *816*, 152491.
[25] L. Meng, H. Chai, Z. Lv, T. Yang, *Opt. Express* **2021**, *29*, 11301.
[26] L. Meng, Z. Lv, H. Chai, X. Yang, T. Yang, *J. Phys. Appl. Phys.* **2022**, *55*, 19LT01.
[27] A. Ohtomo, M. Kawasaki, T. Koida, K. Masubuchi, H. Koinuma, Y. Sakurai, Y. Yoshida, T. Yasuda, Y. Segawa, *Appl. Phys. Lett.* **1998**, *72*, 2466.
[28] T. Makino, Y. Segawa, M. Kawasaki, A. Ohtomo, R. Shiroki, K. Tamura, T. Yasuda, H. Koinuma, *Appl. Phys. Lett.* **2001**, *78*, 1237.
[29] T. Maemoto, N. Ichiba, H. Ishii, S. Sasa, M. Inoue, *J. Phys. Conf. Ser.* **2007**, *59*, 670.
[30] X. Yuan, T. Yamada, L. Meng, *Appl. Phys. Lett.* **2022**, *121*, 152903.
[31] K. Ferri, S. Bachu, W. Zhu, M. Imperatore, J. Hayden, N. Alem, N. Giebink, S. Trolier-McKinstry, J.-P. Maria, *J. Appl. Phys.* **2021**, *130*, 044101.
[32] S. Fichtner, N. Wolff, F. Lofink, L. Kienle, B. Wagner, *J. Appl. Phys.* **2019**, *125*, 114103.
[33] A. Konishi, T. Ogawa, C. A. J. Fisher, A. Kuwabara, T. Shimizu, S. Yasui, M. Itoh, H. Moriwake, *Appl. Phys. Lett.* **2016**, *109*, 102903.
[34] D. A. Scrymgeour, V. Gopalan, T. E. Haynes, *Integr. Ferroelectr.* **2001**, *41*, 35.
[35] D. A. Scrymgeour, A. Sharan, V. Gopalan, K. T. Gahagan, J. L. Casson, R. Sander, J. M. Robinson, F. Muhammad, P. Chandramani, F. Kiamilev, *Appl. Phys. Lett.* **2002**, *81*, 3140.
[36] S. Trolier-McKinstry, R. E. Newnham, *Materials Engineering: Bonding, Structure, and Structure-Property Relationships*, Cambridge University Press, **2018**.
[37] V. Srikant, D. R. Clarke, *J. Appl. Phys.* **1998**, *83*, 5447.
[38] S. Heo, E. Cho, H.-I. Lee, G. S. Park, H. J. Kang, T. Nagatomi, P. Choi, B.-D. Choi, *AIP Adv.* **2015**, *5*, 077167.
[39] J. Tauc, R. Grigorovici, A. Vancu, *Phys. Status Solidi B* **1966**, *15*, 627.
[40] B. D. Viezbicke, S. Patel, B. E. Davis, D. P. Birnie III, *Phys. Status Solidi B* **2015**, *252*, 1700.
[41] S. A. Denev, T. T. A. Lummen, E. Barnes, A. Kumar, V. Gopalan, *J. Am. Ceram. Soc.* **2011**, *94*, 2699.
[42] N. Bloembergen, P. S. Pershan, *Phys. Rev.* **1962**, *128*, 606.
[43] W. N. Herman, L. M. Hayden, *JOSA B* **1995**, *12*, 416.
[44] D. A. Kleinman, *Phys. Rev.* **1962**, *126*, 1977.
[45] R. C. Miller, *Appl. Phys. Lett.* **1964**, *5*, 17.
[46] M. C. Larciprete, M. Centini, *Appl. Phys. Rev.* **2015**, *2*, 031302.

[47] N. Bloembergen, P. S. Pershan, *Phys. Rev.* **1962**, *128*, 606.
[48] W. N. Herman, L. M. Hayden, *JOSA B* **1995**, *12*, 416.
[49] R. Zu, B. Wang, J. He, J.-J. Wang, L. Weber, L.-Q. Chen, V. Gopalan, *Npj Comput. Mater.* **2022**, *8*, 1.
[50] G. Wang, G. T. Kiehne, G. K. L. Wong, J. B. Ketterson, X. Liu, R. P. H. Chang, *Appl. Phys. Lett.* **2002**, *80*, 401.
[51] F. Liang, L. Kang, Z. Lin, Y. Wu, *Cryst. Growth Des.* **2017**, *17*, 2254.
[52] D. N. Nikogosyan, *Nonlinear Optical Crystals: A Complete Survey*, Springer-Verlag, New York, **2005**.
[53] M. Vallade, *Phys. Rev. B* **1975**, *12*, 3755.
[54] D. K. T. Chu, H. Hsiung, *Appl. Phys. Lett.* **1992**, *61*, 1766.
[55] M. E. Lines, A. M. Glass, *Principles and Applications of Ferroelectrics and Related Materials*, Oxford University Press, Oxford, New York, **2001**.
[56] K. Harun, N. Mansor, M. K. Yaakob, M. F. M. Taib, Z. A. Ahmad, A. A. Mohamad, *J. Sol-Gel Sci. Technol.* **2016**, *80*, 56.
[57] D. M. Hoat, V. V. On, D. Khanh Nguyen, M. Naseri, R. Ponce-Pérez, T. V. Vu, J. F. Rivas-Silva, N. N. Hieu, G. H. Cocoletzi, *RSC Adv.* **2020**, *10*, 40411.
[58] D. Damjanovic, *Rep. Prog. Phys.* **1998**, *61*, 1267.
[59] M. J. Haun, E. Furman, S. J. Jang, L. E. Cross, *Ferroelectrics* **1989**, *99*, 13.
[60] D. Damjanovic, *J. Am. Ceram. Soc.* **2005**, *88*, 2663.
[61] M. Akiyama, K. Kano, A. Teshigahara, *Appl. Phys. Lett.* **2009**, *95*, 162107.
[62] T. Zheng, J. Wu, D. Xiao, J. Zhu, *Prog. Mater. Sci.* **2018**, *98*, 552.
[63] M. J. Haun, E. Furman, S. J. Jang, L. E. Cross, *Ferroelectrics* **1989**, *99*, 13.
[64] M. J. Haun, E. Furman, S. J. Jang, L. E. Cross, *Ferroelectrics* **1989**, *99*, 63.
[65] D. Damjanovic, *J. Am. Ceram. Soc.* **2005**, *88*, 2663.
[66] D. Alden, W. Guo, R. Kirste, F. Kaess, I. Bryan, T. Troha, A. Bagal, P. Reddy, L. H. Hernandez-Balderrama, A. Franke, S. Mita, C.-H. Chang, A. Hoffmann, M. Zgonik, R. Collazo, Z. Sitar, *Appl. Phys. Lett.* **2016**, *108*, 261106.
[67] D. Alden, T. Troha, R. Kirste, S. Mita, Q. Guo, A. Hoffmann, M. Zgonik, R. Collazo, Z. Sitar, *Appl. Phys. Lett.* **2019**, *114*, 103504.
[68] J. Park, Y. Yamazaki, M. Iwanaga, H. Jeon, T. Fujiwara, T. Yao, *Opt. Express* **2010**, *18*, 7851.
[69] G. D. Boyd, C. K. N. Patel, *Appl. Phys. Lett.* **1966**, *8*, 313.
[70] P. Giannozzi, S. Baroni, N. Bonini, M. Calandra, R. Car, C. Cavazzoni, D. Ceresoli, G. L. Chiarotti, M. Cococcioni, I. Dabo, A. D. Corso, S. de Gironcoli, S. Fabris, G. Fratesi, R. Gebauer, U. Gerstmann, C. Gougoussis, A. Kokalj, M. Lazzeri, L. Martin-Samos, N. Marzari, F. Mauri, R. Mazzarello, S. Paolini, A. Pasquarello, L. Paulatto, C. Sbraccia, S. Scandolo, G. Sclauzero, A. P. Seitsonen, A. Smogunov, P. Umari, R. M. Wentzcovitch, *J. Phys. Condens. Matter* **2009**, *21*, 395502.
[71] P. Giannozzi, O. Andreussi, T. Brumme, O. Bunau, M. B. Nardelli, M. Calandra, R. Car, C. Cavazzoni, D. Ceresoli, M. Cococcioni, N. Colonna, I. Carnimeo, A. D. Corso, S. de Gironcoli, P. Delugas, R. A. DiStasio, A. Ferretti, A. Floris, G. Fratesi, G. Fugallo, R. Gebauer, U. Gerstmann, F. Giustino, T. Gorni, J. Jia, M. Kawamura, H.-Y. Ko, A. Kokalj, E. Küçükbenli, M. Lazzeri, M. Marsili, N. Marzari, F. Mauri, N. L. Nguyen, H.-V. Nguyen, A. Otero-de-la-Roza, L. Paulatto, S. Poncé, D. Rocca, R. Sabatini, B. Santra, M. Schlipf, A. P. Seitsonen, A.

Smogunov, I. Timrov, T. Thonhauser, P. Umari, N. Vast, X. Wu, S. Baroni, *J. Phys. Condens. Matter* **2017**, *29*, 465901.
[72] M. J. van Setten, M. Giantomassi, E. Bousquet, M. J. Verstraete, D. R. Hamann, X. Gonze, G.-M. Rignanese, *Comput. Phys. Commun.* **2018**, *226*, 39.
[73] K. Lejaeghere, G. Bihlmayer, T. Björkman, P. Blaha, S. Blügel, V. Blum, D. Caliste, I. E. Castelli, S. J. Clark, A. Dal Corso, S. de Gironcoli, T. Deutsch, J. K. Dewhurst, I. Di Marco, C. Draxl, M. Dułak, O. Eriksson, J. A. Flores-Livas, K. F. Garrity, L. Genovese, P. Giannozzi, M. Giantomassi, S. Goedecker, X. Gonze, O. Grånäs, E. K. U. Gross, A. Gulans, F. Gygi, D. R. Hamann, P. J. Hasnip, N. a. W. Holzwarth, D. Iuşan, D. B. Jochym, F. Jollet, D. Jones, G. Kresse, K. Koepernik, E. Küçükbenli, Y. O. Kvashnin, I. L. M. Locht, S. Lubeck, M. Marsman, N. Marzari, U. Nitzsche, L. Nordström, T. Ozaki, L. Paulatto, C. J. Pickard, W. Poelmans, M. I. J. Probert, K. Refson, M. Richter, G.-M. Rignanese, S. Saha, M. Scheffler, M. Schlipf, K. Schwarz, S. Sharma, F. Tavazza, P. Thunström, A. Tkatchenko, M. Torrent, D. Vanderbilt, M. J. van Setten, V. Van Speybroeck, J. M. Wills, J. R. Yates, G.-X. Zhang, S. Cottenier, *Science* **2016**, *351*, aad3000.
[74] D. R. Hamann, *Phys. Rev. B* **2013**, *88*, 085117.
[75] J. P. Perdew, J. A. Chevary, S. H. Vosko, K. A. Jackson, M. R. Pederson, D. J. Singh, C. Fiolhais, *Phys. Rev. B* **1992**, *46*, 6671.
[76] A. D. Becke, *Phys. Rev. A* **1988**, *38*, 3098.
[77] D. C. Langreth, M. J. Mehl, *Phys. Rev. B* **1983**, *28*, 1809.
[78] J. P. Perdew, K. Burke, M. Ernzerhof, *Phys. Rev. Lett.* **1996**, *77*, 3865.
[79] A. H. Larsen, J. J. Mortensen, J. Blomqvist, I. E. Castelli, R. Christensen, M. Dułak, J. Friis, M. N. Groves, B. Hammer, C. Hargus, E. D. Hermes, P. C. Jennings, P. B. Jensen, J. Kermode, J. R. Kitchin, E. L. Kolsbjerg, J. Kubal, K. Kaasbjerg, S. Lysgaard, J. B. Maronsson, T. Maxson, T. Olsen, L. Pastewka, A. Peterson, C. Rostgaard, J. Schiøtz, O. Schütt, M. Strange, K. S. Thygesen, T. Vegge, L. Vilhelmsen, M. Walter, Z. Zeng, K. W. Jacobsen, *J. Phys. Condens. Matter* **2017**, *29*, 273002.
[80] I. Timrov, N. Marzari, M. Cococcioni, *Phys. Rev. B* **2018**, *98*, 085127.
[81] I. Timrov, N. Marzari, M. Cococcioni, *Comput. Phys. Commun.* **2022**, *279*, 108455.
[82] N. E. Kirchner-Hall, W. Zhao, Y. Xiong, I. Timrov, I. Dabo, *Appl. Sci.* **2021**, *11*, 2395.
[83] S. Baroni, S. de Gironcoli, A. Dal Corso, P. Giannozzi, *Rev. Mod. Phys.* **2001**, *73*, 515.
[84] R. Resta, D. Vanderbilt, in *Phys. Ferroelectr. Mod. Perspect.*, Springer, Berlin, Heidelberg, **2007**, pp. 31–68.
[85] N. A. Spaldin, *J. Solid State Chem.* **2012**, *195*, 2.
[86] S. Jesse, S. V. Kalinin, R. Proksch, A. P. Baddorf, B. J. Rodriguez, *Nanotechnology* **2007**, *18*, 435503.

**Supporting Information**

**Large Enhancements in Optical and Piezoelectric Properties in Ferroelectric $Zn_{1-x}Mg_xO$ Thin Films through Engineering Electronic and Ionic Anharmonicities**

*Rui Zu, Gyunghyun Ryu, Kyle P. Kelley, Steven M. Baksa, Leonard C Jacques, Bo Wang, Kevin Ferri, Jingyang He, Long-Qing Chen, Ismaila Dabo, Susan Trolier-McKinstry, Jon-Paul Maria, Venkatraman Gopalan**

## 1. AFM and SEM of $Zn_{1-x}Mg_xO$

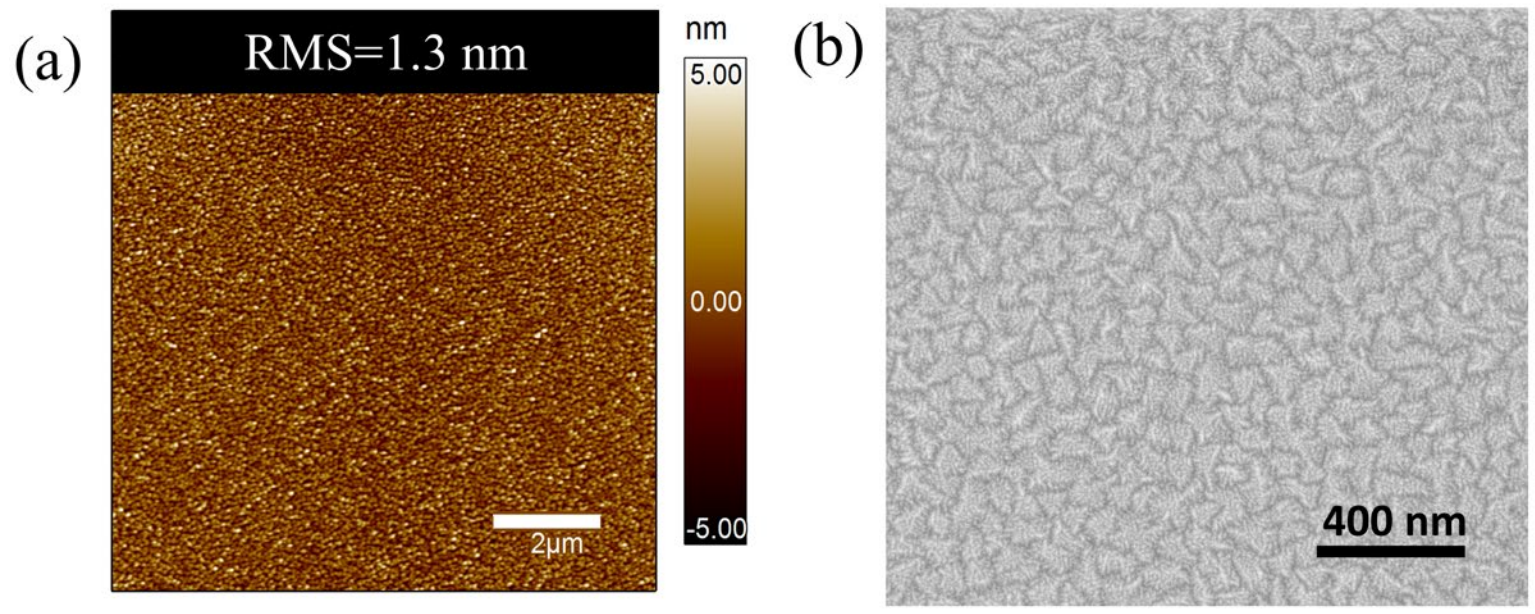

**Figure S1**. (a) AFM of $Zn_{0.72}Mg_{0.28}$ with a surface roughness RMS = 1.3nm (b) SEM of $Zn_{0.72}Mg_{0.28}$.

## 2. Refractive Indices of $Zn_{1-x}Mg_xO$

Cauchy dispersion relations take the form of $n_i(\lambda) = A_i + \frac{B_i}{\lambda^2} + \frac{C_i}{\lambda^4}$, where the subscript $i$ denotes the birefringence (o for ordinary, e for extraordinary). Their parameters are summarized in Table S1.

**Table S1.** Cauchy parameters of $Zn_{1-x}Mg_xO$ films. *A*, *B,* and *C* are Cauchy parameters. Subscripts o and e represent ordinary and extraordinary parts.

| Mg Concentration, *x* | $A_o$ | $B_o$ | $C_o$ | $A_e$ | $B_e$ | $C_e$ |
|---|---|---|---|---|---|---|
| 0 | 1.928 | 0.01068 | 0.00530 | 1.848 | 0.08447 | −0.00818 |
| 0.16 | 1.841 | 0.02244 | 0.00109 | 1.780 | 0.05220 | −0.00329 |
| 0.23 | 1.822 | 0.02398 | 0.00064 | 1.817 | 0.01516 | 0.00170 |
| 0.28 | 1.805 | 0.02070 | 0.00072 | 1.794 | 0.02008 | 0.00056 |
| 0.36 | 1.785 | 0.02050 | 0.00022 | 1.775 | 0.01883 | 0.00063 |

## 3. Power-dependent SHG of $Zn_{0.63}Mg_{0.37}O$

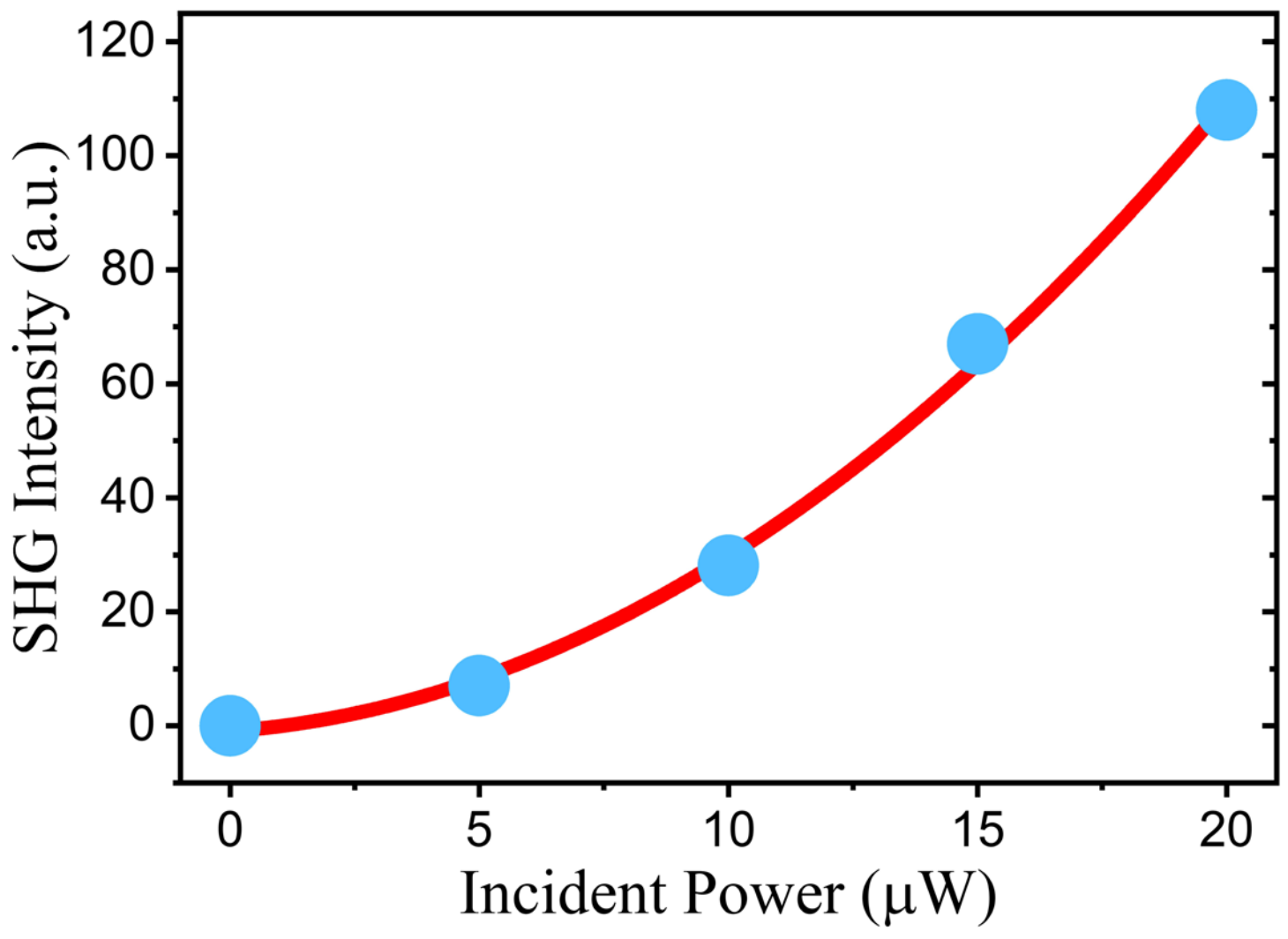

**Figure S2**. Power-dependent SHG of $Zn_{0.63}Mg_{0.37}O$. The blue dots are experimental results, and the red line is the quadratic fitting.

## 4. SHG of ZnO Single Crystal Using ♯SHAARP

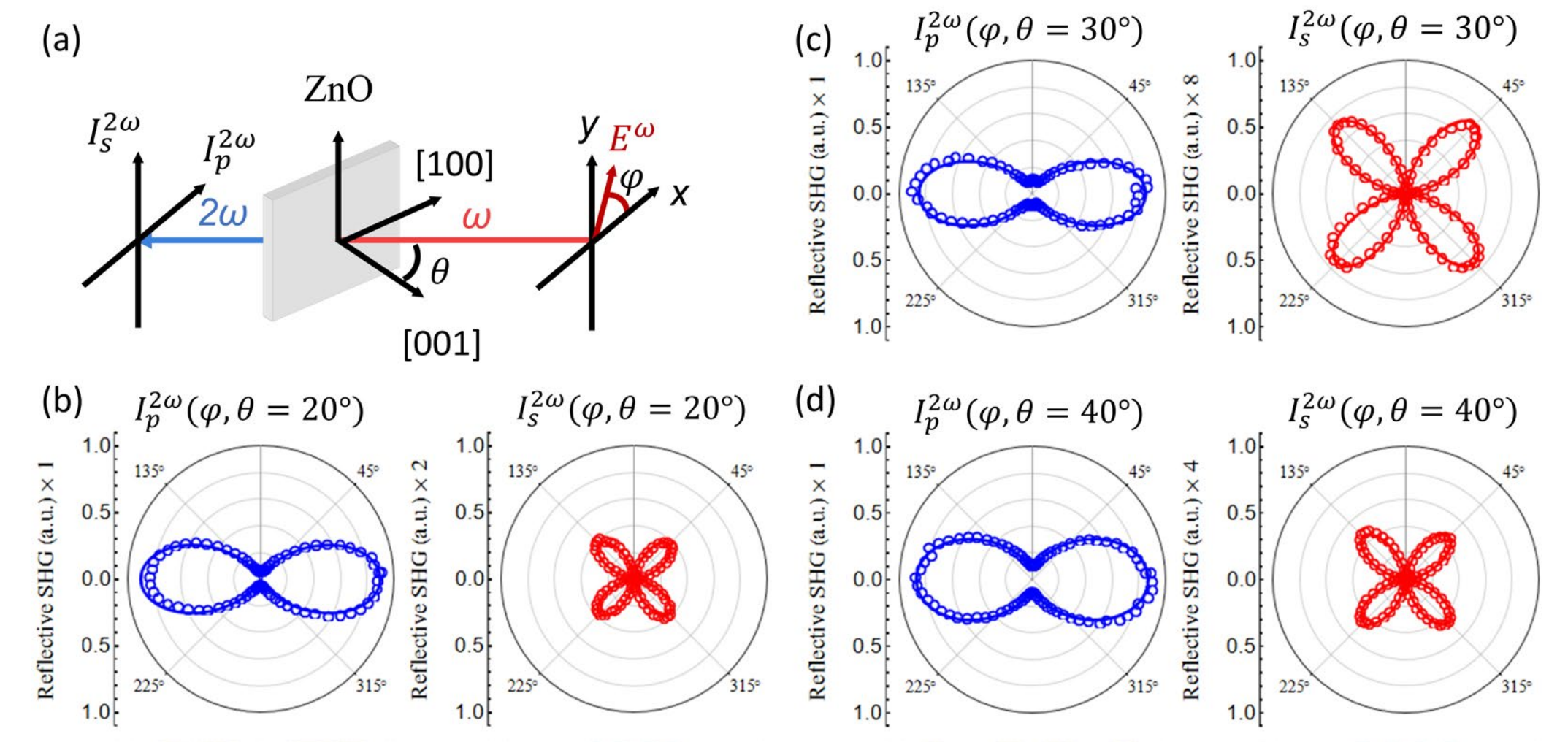

**Figure S3**. SHG polarimetry of 0.5mm thick ZnO (0001) single crystal. (a) Measurement geometry. Blue and red are fundamental light and SHG beam, respectively. $\varphi$ is the azimuthal angle for the polarimetry. (b-d) SHG polar plots measured at different incident angles. (b) $\theta = 20°$, (c) $\theta = 30°$, (d) $\theta = 40°$. Blue is *p*- polarized, and red is *s*- polarized SHG.

## 5. Summary of nonlinear optical susceptibilities of ZnO single crystal and $Zn_{1-x}Mg_xO$ thin film

**Table S2**. Nonlinear optical susceptibility (pm $V^{-1}$) of $Zn_{1-x}Mg_xO$ films and ZnO single crystal (SC).

| *d* | *x* = 0 | 0.16 | 0.23 | 0.28 | 0.37 | SC |
|---|---|---|---|---|---|---|
| $d_{33}$ | $7.05 \pm 2.81$ | $10.07 \pm 2.32$ | $10.08 \pm 2.68$ | $7.86 \pm 2.71$ | $5.82 \pm 2.30$ | $7.30 \pm 1.15$ |
| $d_{31}$ | $-0.92 \pm 0.23$ | $-0.81 \pm 0.21$ | $-0.77 \pm 0.26$ | $-0.54 \pm 0.30$ | $-0.53 \pm 0.24$ | $-0.95 \pm 0.17$ |
| $d_{15}$ | $-1.37 \pm 0.07$ | $-1.22 \pm 0.06$ | $-0.99 \pm 0.06$ | $-0.96 \pm 0.06$ | $-0.83 \pm 0.06$ | $-0.61 \pm 0.10$ |

## 6. Band Excitation PFM (BEPFM) of $Zn_{1-x}Mg_xO$

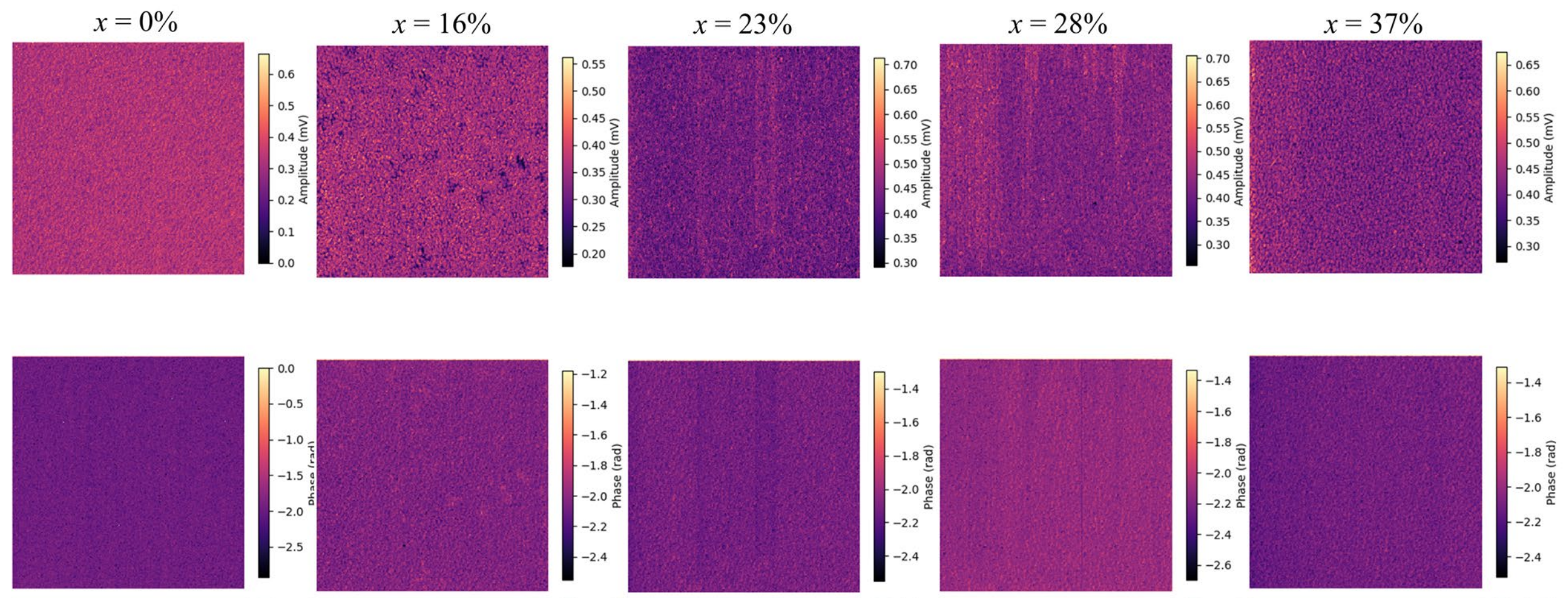

**Figure S4**. Band excitation PFM (PFM) of all concentrations in $Zn_{1-x}Mg_xO$.

## 7. Landau-Ginzburg-Devonshire (LGD) Theory for Ionic Response

Landau-Ginzburg-Devonshire (LGD) theory has been widely applied in explaining ferroelectricity and coupling phenomena related to polarization.[1–3] The Gibbs free energy can be written in the form indicated below,

$$G(P) = -a_i P^2 + \beta_i P^4 + \frac{1}{2} S X^2 - Q X P^2 \cdots \tag{S1}$$

where $U$, $P$, $S$, $Q$, $X$, and $\alpha, \beta$ are the free energy potential, polarization, compliance, electrostriction, stress, and Landau expansion coefficients. The subscript $i$ represents the contributions from ions. Since higher-order terms tend to play less effect on the energy, for the scope of the study, we keep the form up to fourth-order term $P^4$. The equilibrium state refers to the polarization state with minimum energy ($G$), and the corresponding polarization refers to the spontaneous polarization ($P_{s,i}$) of the system. Based on the expression of the energy potential well, five key parameters, namely, polarization ($P_{s,i}$), energy barrier ($U_b$), dielectric susceptibility ($\chi_i$), anharmonicity ($A_i$), and piezoelectric coefficient ($d^{\mathrm{Piezo}}$) can be obtained following Equation S1. Under zero stress, the magnitude of the spontaneous polarization ($P_{s,i}$) can be obtained following $\frac{\partial G}{\partial P} = 0$. The energy barrier ($U_b$) represents the energy difference between $G(0)$ and $G(P_{s,i})$, which measures the energy barrier to overcome during polarization switching. The dielectric susceptibility can be obtained from $\chi_i = \varepsilon_0^{-1} \left(\frac{\partial^2 G}{\partial P^2}\right)^{-1}_{P=P_{s,i}}$. Taking the factor of $P^3$ using Taylor expansion at $P_{s,i}$ can reveal the anharmonicity ($A_i$) of the potential well. The origin of piezoelectricity using double well potential has been laid out by Haun and Damjanovic[4,5,1,6]. Therefore, five properties as a function of Landau expansion coefficients can be summarized below[4],

$$P_{s,i} = \sqrt{\frac{\alpha_i}{2\beta_i}} \tag{S2}$$

$$U_b = \frac{\alpha_i^2}{4\beta_i} \tag{S3}$$

$$\chi_i = \frac{1}{4\alpha_i \varepsilon_0} \tag{S4}$$

$$A_i = 2\sqrt{2\alpha_i \beta_i} \tag{S5}$$

$$d^{\mathrm{Piezo}} = \frac{Q}{2\varepsilon_0\sqrt{2\alpha_i\beta_i}} = \frac{Q}{\varepsilon_0 A_i} \tag{S6}$$

where $\varepsilon_0$ represents the vacuum permittivity.

**Figure S5** outlines the ionic potential wells and five properties as a function of Landau coefficients. Purple arrows highlight the pathway as increasing Mg concentration. **Figure S5(a)** illustrates the ionic double-well potential for $Zn_{1-x}Mg_xO$. The two wells with opposite polarizations indicate a switchable polarization, and the equal energy for both wells suggests that the two states are at equilibrium and energetically favorable. Landau's theory provides built-in systematic relations for physical properties and can be used effectively to understand, predict and design materials' properties. As increasing Mg concentration, the potential well changes from blue to yellow, as indicated by the purple arrows, leading to ionic properties changes in the following discussions. Figures S5(b-f) depict spontaneous polarization $(P_{s,i})$, energy barrier $(U_b)$, dielectric susceptibility $(\chi_i)$, anharmonicity $(A_i)$, and piezoelectric coefficient $(d^{\text{Piezo}})$ following Equations S2-6 as a function of Landau coefficients $(\alpha_i, \beta_i)$. As indicated by the purple arrows, Figures S5(b) and (c) show that both $P_{s,i}$ and $U_b$ will decrease consistently with higher Mg concentration. Both DFT and experiment[7] confirms that $P_{s,i}$ reduces as increasing Mg concentration. Moreover, experimental observation[7] reveals the emerging ferroelectricity as substituting more Zn with Mg, indicating the reduction of the energy barrier. Figure S5(d) predicts the enhancement of the dielectric constant, which is also well supported by the experimental results[7]. This agreement highlights that $\alpha_i$ is decreasing in the $Zn_{1-x}Mg_xO$ system. On the other hand, $\beta_i$ is expected to increase in the $Zn_{1-x}Mg_xO$, since it is the measure of the slope of outer energy wall and describes the repulsive energy as Zn ions move toward the oxygen atom along the *c* axis. Since the *c* axis shrinks with more Mg, the closer distance between Zn(Mg) and O along the *c* axis indicates higher repulsive energy, leading to a higher $\beta_i$. Figure S5(e) and (f) further illustrate changes of $A_i$ and $d^{\text{Piezo}}$ in $Zn_{1-x}Mg_xO$. In particular, the inverse proportionality between $A_i$ and $d^{\text{Piezo}}$ emphasizes that a less unsymmetric potential well at $P_{s,i}$ is essential for a large $d^{\text{Piezo}}$, in contrast to the $d^{\text{SHG}}$ where an increase in the electronic nonlinearity is required. However, utilizing anharmonicity at distinct frequencies lays out a possible route for achieving enhancements of both SHG and piezoelectricity simultaneously.

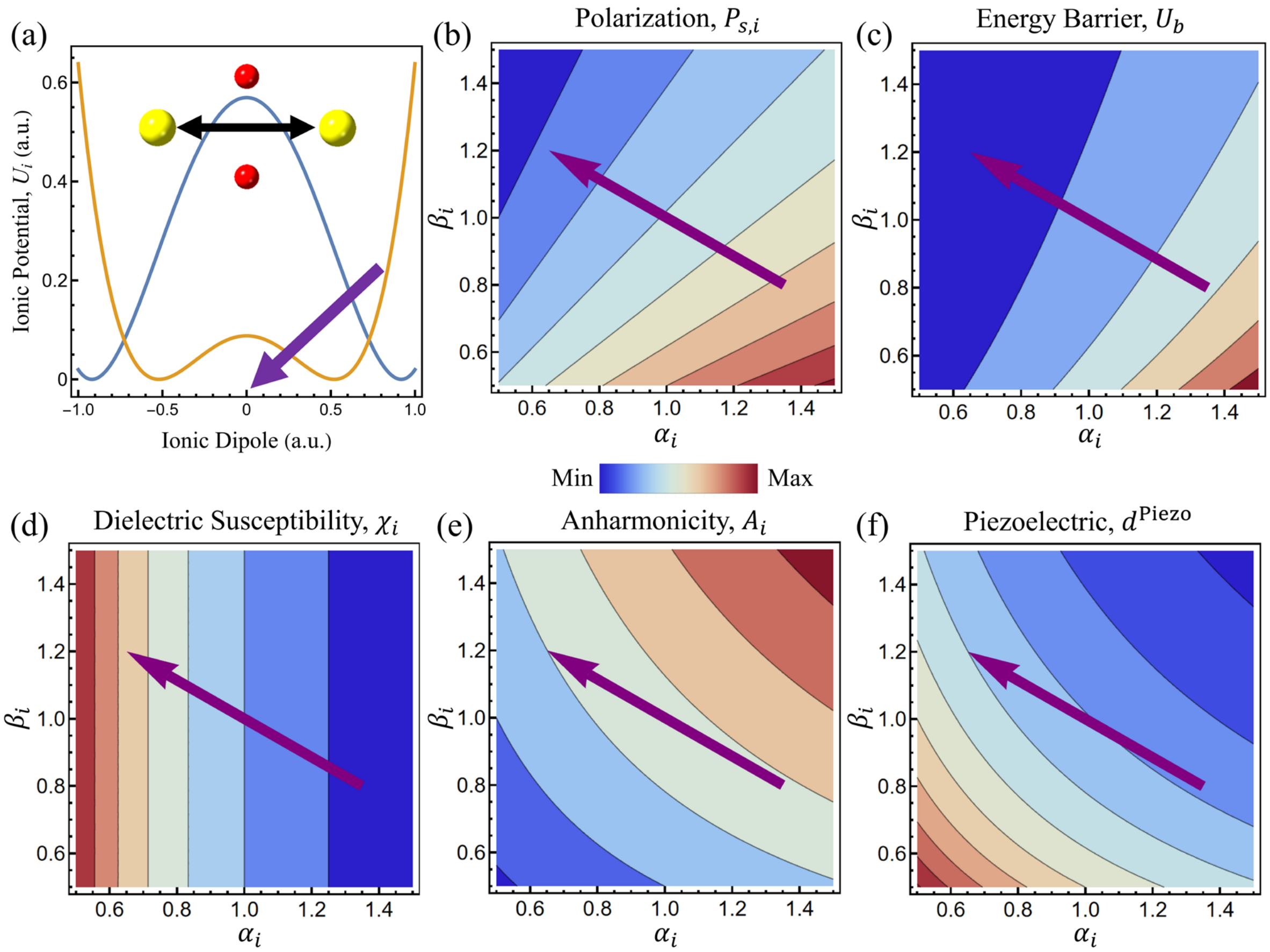

**Figure S5**. Ionic tdouble-well potential and polarization-related properties as a function of Landau coefficients ($\alpha_i, \beta_i$). Purple arrows highlight the changes as increasing Mg concentration. (a) Ionic double-well potential of $Zn_{1-x}Mg_xO$. The inset represents the ferroelectric ionic displacement. Red atoms are oxygen, and yellow atoms are Zn(Mg). The Blue and yellow potentials are pure ZnO and $Zn_{1-x}Mg_xO$, respectively. (b-f) Contour plots of polarization-related properties covering (b) spontaneous polarization ($P_{s,i}$), (c) energy barrier ($U_b$), (d) dielectric susceptibility ($\chi_i$), (e) anharmonicity ($A_i$), (f) piezoelectric coefficient ($d^{\text{Piezo}}$).

## 8. SHG Fitting Analysis

In the SHG polarimetry analysis, three incident polarization angles are representative for uniquely determining nonlinear optical coefficients. Semi-analytical solutions were obtained from

♯SHAARP based on multi-reflection calculation in the slab waveguide, where both forward and backward propagating waves are considered for both linear and nonlinear frequencies.[8–10] The measurement was performed at 45-degree reflection geometry. The measured SHG intensity was then compared with a reference crystal, a wedged X-cut $LiNbO_3$ in this study, to calibrate the absolute nonlinear optical coefficients. Both films and $LiNbO_3$ are aligned such that the optical axes are in the plane of incidence. The $d_{33}^{\mathrm{SHG}}$ of $LiNbO_3$ was 19.5 pm/V, calibrated using Miller's rule from Shoji et al.'s work,[9,11] and the SHG expression of wedged X-cut $LiNbO_3$ can be found from our previous work.[12] The thicknesses of films are set to 150 nm, as confirmed by ellipsometry.

For $Zn_{1-x}Mg_xO$ films, the semi-analytical expressions are shown below,

$$I_p^{2\omega}(x=0,\varphi=0) = 0.0915(d_{15}^{\mathrm{SHG}})^2 + 0.0393 d_{15}^{\mathrm{SHG}} d_{31}^{\mathrm{SHG}} + 0.0043(d_{31}^{\mathrm{SHG}})^2 + 0.0061 d_{15}^{\mathrm{SHG}} d_{33}^{\mathrm{SHG}} + 0.0013 d_{31}^{\mathrm{SHG}} d_{33}^{\mathrm{SHG}} + 0.000103(d_{33}^{\mathrm{SHG}})^2 \quad \text{(S7)}$$

$$I_p^{2\omega}\left(x=0,\varphi=\frac{\pi}{4}\right) = 0.0229(d_{15}^{\mathrm{SHG}})^2 + 0.0211 d_{15}^{\mathrm{SHG}} d_{31}^{\mathrm{SHG}} + 0.0056(d_{31}^{\mathrm{SHG}})^2 + 0.0015 d_{15}^{\mathrm{SHG}} d_{33}^{\mathrm{SHG}} + 0.0007 d_{31}^{\mathrm{SHG}} d_{33}^{\mathrm{SHG}} + 0.000026(d_{33}^{\mathrm{SHG}})^2 \quad \text{(S8)}$$

$$I_p^{2\omega}\left(x=0,\varphi=\frac{\pi}{2}\right) = 0.0079(d_{31}^{\mathrm{SHG}})^2 \quad \text{(S9)}$$

$$I_s^{2\omega}\left(x=0,\varphi=\frac{\pi}{4}\right) = 0.0263(d_{15}^{\mathrm{SHG}})^2 \quad \text{(S10)}$$

$$I_p^{2\omega}(x=0.16,\varphi=0) = 0.1123(d_{15}^{\mathrm{SHG}})^2 + 0.0481 d_{15}^{\mathrm{SHG}} d_{31}^{\mathrm{SHG}} + 0.0052(d_{31}^{\mathrm{SHG}})^2 + 0.0080 d_{15}^{\mathrm{SHG}} d_{33}^{\mathrm{SHG}} + 0.0017 d_{31}^{\mathrm{SHG}} d_{33}^{\mathrm{SHG}} + 0.000143(d_{33}^{\mathrm{SHG}})^2 \quad \text{(S11)}$$

$$I_p^{2\omega}\left(x=0.16,\varphi=\frac{\pi}{4}\right) = 0.0281(d_{15}^{\mathrm{SHG}})^2 + 0.0259 d_{15}^{\mathrm{SHG}} d_{31}^{\mathrm{SHG}} + 0.0069(d_{31}^{\mathrm{SHG}})^2 + 0.0020 d_{15}^{\mathrm{SHG}} d_{33}^{\mathrm{SHG}} + 0.0010 d_{31}^{\mathrm{SHG}} d_{33}^{\mathrm{SHG}} + 0.000036(d_{33}^{\mathrm{SHG}})^2 \quad \text{(S12)}$$

$$I_p^{2\omega}\left(x=0.16,\varphi=\frac{\pi}{2}\right) = 0.00980(d_{31}^{\mathrm{SHG}})^2 \quad \text{(S13)}$$

$$I_s^{2\omega}\left(x=0.16,\varphi=\frac{\pi}{4}\right)=0.0338(d_{15}^{\mathrm{SHG}})^2 \tag{S14}$$

$$I_p^{2\omega}(x=0.23,\varphi=0)=0.1188(d_{15}^{\mathrm{SHG}})^2+0.0479d_{15}^{\mathrm{SHG}}d_{31}^{\mathrm{SHG}}+0.0050(d_{31}^{\mathrm{SHG}})^2+0.0082d_{15}^{\mathrm{SHG}}d_{33}^{\mathrm{SHG}}+$$

$$0.0017d_{31}^{\mathrm{SHG}}d_{33}^{\mathrm{SHG}}+0.000147(d_{33}^{\mathrm{SHG}})^2 \tag{S15}$$

$$I_p^{2\omega}\left(x=0.23,\varphi=\frac{\pi}{4}\right)=0.0297(d_{15}^{\mathrm{SHG}})^2+0.0247d_{15}^{\mathrm{SHG}}d_{31}^{\mathrm{SHG}}+0.0062(d_{31}^{\mathrm{SHG}})^2+0.0021d_{15}^{\mathrm{SHG}}d_{33}^{\mathrm{SHG}}+$$

$$0.0009d_{31}^{\mathrm{SHG}}d_{33}^{\mathrm{SHG}}+0.000037(d_{33}^{\mathrm{SHG}})^2 \tag{S16}$$

$$I_p^{2\omega}\left(x=0.23,\varphi=\frac{\pi}{2}\right)=0.0085(d_{31}^{\mathrm{SHG}})^2 \tag{S17}$$

$$I_s^{2\omega}\left(x=0.23,\varphi=\frac{\pi}{4}\right)=0.0370(d_{15}^{\mathrm{SHG}})^2 \tag{S18}$$

$$I_p^{2\omega}(x=0.28,\varphi=0)=0.1274(d_{15}^{\mathrm{SHG}})^2+0.0512d_{15}^{\mathrm{SHG}}d_{31}^{\mathrm{SHG}}+0.0053(d_{31}^{\mathrm{SHG}})^2+0.0090d_{15}^{\mathrm{SHG}}d_{33}^{\mathrm{SHG}}+$$

$$0.0019d_{31}^{\mathrm{SHG}}d_{33}^{\mathrm{SHG}}+0.000165(d_{33}^{\mathrm{SHG}})^2 \tag{S19}$$

$$I_p^{2\omega}\left(x=0.28,\varphi=\frac{\pi}{4}\right)=0.0319(d_{15}^{\mathrm{SHG}})^2+0.0263d_{15}^{\mathrm{SHG}}d_{31}^{\mathrm{SHG}}+0.0066(d_{31}^{\mathrm{SHG}})^2+0.0023d_{15}^{\mathrm{SHG}}d_{33}^{\mathrm{SHG}}+$$

$$0.0010d_{31}^{\mathrm{SHG}}d_{33}^{\mathrm{SHG}}+0.000041(d_{33}^{\mathrm{SHG}})^2 \tag{S20}$$

$$I_p^{2\omega}\left(x=0.28,\varphi=\frac{\pi}{2}\right)=0.0090(d_{31}^{\mathrm{SHG}})^2 \tag{S21}$$

$$I_s^{2\omega}\left(x=0.28,\varphi=\frac{\pi}{4}\right)=0.0407(d_{15}^{\mathrm{SHG}})^2 \tag{S22}$$

$$I_p^{2\omega}(x=0.36,\varphi=0)=0.1442(d_{15}^{\mathrm{SHG}})^2+0.0605d_{15}^{\mathrm{SHG}}d_{31}^{\mathrm{SHG}}+0.0065(d_{31}^{\mathrm{SHG}})^2+0.0110d_{15}^{\mathrm{SHG}}d_{33}^{\mathrm{SHG}}+$$

$$0.0023d_{31}^{\mathrm{SHG}}d_{33}^{\mathrm{SHG}}+0.000212(d_{33}^{\mathrm{SHG}})^2 \tag{S23}$$

$$I_p^{2\omega}\left(x=0.36,\varphi=\frac{\pi}{4}\right)=0.0360(d_{15}^{\mathrm{SHG}})^2+0.0323d_{15}^{\mathrm{SHG}}d_{31}^{\mathrm{SHG}}+0.0086(d_{31}^{\mathrm{SHG}})^2+0.0027d_{15}^{\mathrm{SHG}}d_{33}^{\mathrm{SHG}}+$$

$$0.0013d_{31}^{\mathrm{SHG}}d_{33}^{\mathrm{SHG}}+0.000053(d_{33}^{\mathrm{SHG}})^2 \tag{S24}$$

$$I_p^{2\omega}\left(x=0.36,\varphi=\frac{\pi}{2}\right)=0.0121(d_{31}^{\mathrm{SHG}})^2 \tag{S25}$$

$$I_s^{2\omega}\left(x=0.36,\varphi=\frac{\pi}{4}\right)=0.0464(d_{15}^{\mathrm{SHG}})^2 \tag{S26}$$

where $x$ and $\varphi$ stand for concentration and incident polarization, respectively. Subscript $p$ and $s$ represent $p$- or $s$- polarized light. Since there will be limited electric fields projected along the $d_{33}^{\mathrm{SHG}}$ direction, at 45-degree incidence, only a small portion of $d_{33}^{\mathrm{SHG}}$ can be probed. Therefore, $I_p^{2\omega}(\varphi)$ is highly sensitive to the variation of both $d_{15}^{\mathrm{SHG}}$ and $d_{31}^{\mathrm{SHG}}$. Commonly, Kleiman's symmetry is assumed where $d_{15}^{\mathrm{SHG}} = d_{31}^{\mathrm{SHG}}$, influencing the final value of $d_{33}^{\mathrm{SHG}}$. With our semi-analytical expression, we can impose such assumption in the analysis, which results in a $d_{33}^{\mathrm{SHG}}$ of nearly 50 pm/V for x=0.28 and 0.36, as shown in **Figure. S6**. Nonetheless, the enhancement of nonlinear susceptibilities can be observed both with and without Kleiman's symmetry.

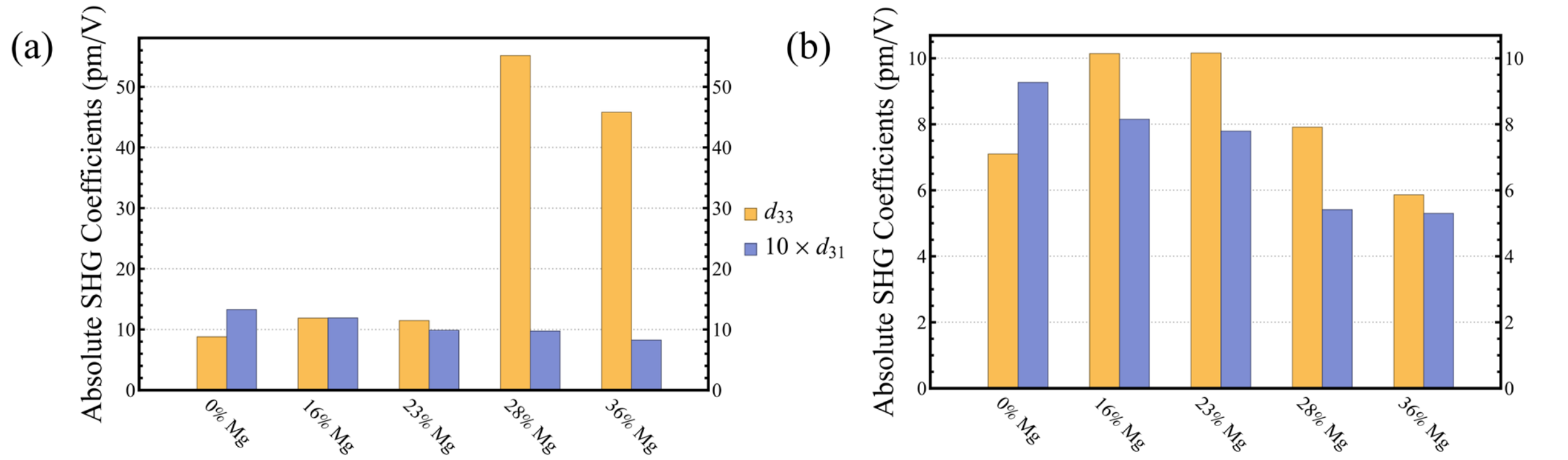

Figure S6. Absolute SHG coefficient of $Zn_{1-x}Mg_xO$. (a) Applying the erroneous assumption of Kleiman's symmetry ($d_{15}^{\mathrm{SHG}} = d_{31}^{\mathrm{SHG}}$). This could be one possible reason for the large enhancement reported previously.[13] (b) Fitting results without Kleiman's symmetry.

## 9. DFT+*U* methodology for $Zn_{1-x}Mg_xO$ calculations

At variance to empirical methods, linear response theory enables one to calculate the Hubbard $U$ parameter as an intrinsic property of individual atomic sites in a given material. In this work, density-functional perturbation theory and linear-response approximations are used to predict these parameters.[14,15] Here, a brief outline of the methodology is provided.

In DFT+$U$, the energy is split into two terms: the DFT energy and the correction from the Hubbard $U$ parameters.

$$E_{\mathrm{DFT}+U} = E_{\mathrm{DFT}} + E_U \tag{S27}$$

$$E_U = \frac{1}{2}\sum_{I\sigma m_1 m_2} U^I\left(\delta_{m_1 m_2} - n^{I\sigma}_{m_1 m_2}\right)n^{I\sigma}_{m_2 m_1}, \tag{S28}$$

where $I$ is the atomic site index, $\sigma$ is the spin label, and $m$ is the magnetic number. The atomic occupation matrices $n^{I\sigma}_{m_1 m_2}$ are constructed by projecting the Kohn-sham wavefunctions onto the Hubbard manifold using Löwdin orthogonalization.[15] Using density-functional perturbation theory and linear-response approximations, a Hubbard $U$ matrix is constructed as the difference between the bare and screened inverse susceptibilities:

$$U^I = (\chi_0^{-1} - \chi^{-1})_{II}. \tag{S29}$$

These susceptibilities can be formulated as a response of the atomic occupations to a potential shift to the Hubbard manifold:

$$\chi_{ij} = \sum_{\sigma m}^{\sigma' m'} \frac{dn^{I\sigma}_{mm}}{d\alpha^J} \tag{S30}$$

where $\alpha^J$ is the magnitude of the perturbation on $J$ atomic site. Although Kirchner-Hall *et al.* conducted a benchmarking analysis of binary oxides of transition metals and p-block elements with $d^{10}$ electronic configurations, there does not exist guidance on Hubbard $U$ parameters for binary oxides of $d^{10}$ transition metals (i.e., Zn).[15] Therefore, a benchmarking analysis was conducted for wurtzite ZnO in **Table S3**. Fixed-cell and relaxed-cell approaches were considered in the analysis, and Hubbard $U$ parameters for Zn-3*d* and O-2*p* orbitals were calculated since these orbitals dominate the valence band edge. The analysis showed that GGA+$U$(O-2p) using a fixed cell was in reasonable agreement with experiment at moderate computational cost. It should be noted that the Hubbard *U* parameters depend not only on the atomic orbitals, but also on the local environment of the orbitals. Therefore, the average $U_O$ parameters and its range are reported for $x_{Mg}$ = 0-40%, as well as its corresponding band gap in **Table S4**.

**Table S3**: Comparative assessment of electronic-structure calculations for ZnO. GGA+*U*(O-2p). with fixed (PBE) geometry with $U_O$ = 10.24 eV is found to be in closest agreement with experiment.

| Hubbard Method | Hubbard *U* Parameters (eV) | Lattice Parameters (Å) | Band Gap (eV) |
|---|---|---|---|
| GGA+*U*(O-2p) with fixed geometry | $U_O$ = 10.24 | *a* = 3.28, *c* = 5.28 | 3.29 |
| GGA+*U*(O-2p) with optimized geometry | $U_O$ = 17.11 | *a* = 3.20, *c* = 5.12 | 4.61 |
| PBE | — | *a* = 3.28, *c* = 5.28 | 0.81 |
| SCAN | — | — | 1.14 |
| HSE06 | — | — | 2.41 |
| Experiment[1,2] | — | *a* = 3.250, *c* = 5.206 | 3.37 |

**Table S4**: Calculated Hubbard *U* parameters of $Zn_{1-x}Mg_xO$ and its corresponding Hubbard-corrected band gap. Here, the GGA+*U*(O-2*p*) with fixed (PBE) geometry was used to calculate the Hubbard *U* parameters. The average value and range of the Hubbard *U* parameters are reported.

| $x_{Mg}$ (%) | Average $U_O$ (eV) | Range of $U_O$ (eV) | Band Gap (eV) |
|---|---|---|---|
| 0.0 | 10.27 | -- | 3.29 |
| 5.6 | 10.60 | 10.3-11.6 | 3.48 |
| 11.1 | 10.58 | 10.3-11.6 | 3.59 |
| 16.7 | 10.77 | 10.3-11.7 | 3.74 |
| 22.2 | 10.94 | 10.3-12.5 | 3.87 |
| 27.8 | 11.14 | 10.3-12.6 | 4.00 |
| 33.3 | 11.32 | 10.4-13.7 | 4.14 |
| 38.9 | 11.54 | 10.5-13.8 | 4.24 |

# Reference

[1] D. Damjanovic, *Rep. Prog. Phys.* **1998**, *61*, 1267.
[2] M. J. Haun, E. Furman, S. J. Jang, L. E. Cross, *Ferroelectrics* **1989**, *99*, 13.
[3] D. Damjanovic, *J. Am. Ceram. Soc.* **2005**, *88*, 2663.
[4] M. J. Haun, E. Furman, S. J. Jang, L. E. Cross, *Ferroelectrics* **1989**, *99*, 13.
[5] M. J. Haun, E. Furman, S. J. Jang, L. E. Cross, *Ferroelectrics* **1989**, *99*, 63.
[6] D. Damjanovic, *J. Am. Ceram. Soc.* **2005**, *88*, 2663.
[7] K. Ferri, S. Bachu, W. Zhu, M. Imperatore, J. Hayden, N. Alem, N. Giebink, S. Trolier-McKinstry, J.-P. Maria, *J. Appl. Phys.* **2021**, *130*, 044101.
[8] C. R. Pollock, *Fundamentals of Optoelectronics*, Irwin, **1995**.
[9] I. Shoji, T. Kondo, A. Kitamoto, M. Shirane, R. Ito, *JOSA B* **1997**, *14*, 2268.
[10] N. Bloembergen, P. S. Pershan, *Phys. Rev.* **1962**, *128*, 606.
[11] R. W. Boyd, D. Prato, *Nonlinear Optics*, Academic Press, Amsterdam ; Boston, **2008**.
[12] R. Zu, B. Wang, J. He, J.-J. Wang, L. Weber, L.-Q. Chen, V. Gopalan, *Npj Comput. Mater.* **2022**, *8*, 1.
[13] L. Meng, H. Chai, Z. Lv, T. Yang, *Opt. Express* **2021**, *29*, 11301.
[14] I. Timrov, N. Marzari, M. Cococcioni, *Phys. Rev. B* **2018**, *98*, 085127.
[15] N. E. Kirchner-Hall, W. Zhao, Y. Xiong, I. Timrov, I. Dabo, *Appl. Sci.* **2021**, *11*, 2395.